\documentclass[twocolumn, switch]{article} 

\usepackage{arxiv-2col}

\usepackage[utf8]{inputenc} 
\usepackage[T1]{fontenc}   
\usepackage{hyperref}       
\usepackage{url}           
\usepackage{booktabs}       
\usepackage{amsfonts}       
\usepackage{nicefrac}       
\usepackage{microtype}      
\usepackage{lipsum}		    
\usepackage{graphicx}
\usepackage{siunitx}
\usepackage{hyperref}
\hypersetup{
    	colorlinks=true,    
        linkcolor=red,  
        urlcolor=red,
        citecolor=green
}

\usepackage[numbers,square,sort&compress]{natbib}
\usepackage{subfig}				
\usepackage{authblk}
\usepackage{xspace}
\usepackage{url}
\usepackage{amsfonts,amsmath}
\usepackage{xcolor}
\usepackage[flushleft]{threeparttable}
\definecolor{reviewcolor}{rgb}{0,0,0}
\definecolor{newaddcolor}{rgb}{0,0,0}

\definecolor{newcolor}{rgb}{.8,.349,.1}
\newcommand{\ie}{\emph{i.e.}\xspace}
\newcommand{\eg}{\emph{e.g.}\xspace}

\newcommand{\vct}[1]{\boldsymbol{#1}} 
\newcommand{\mat}[1]{\boldsymbol{#1}}

\usepackage{titlesec}
\titlespacing\section{0pt}{12pt plus 3pt minus 3pt}{1pt plus 1pt minus 1pt}
\titlespacing\subsection{0pt}{10pt plus 3pt minus 3pt}{1pt plus 1pt minus 1pt}
\titlespacing\subsubsection{0pt}{8pt plus 3pt minus 3pt}{1pt plus 1pt minus 1pt}

\title{Impact of loss functions on the performance of a deep neural network designed to restore low-dose digital mammography}

\author[1,4,5,6]{Hongming Shan}
\author[2,6]{Rodrigo B. Vimieiro}
\author[2,3]{Lucas R. Borges}
\author[2,7]{Marcelo A. C. Vieira}
\author[1,7]{Ge Wang}

\affil[1]{Dept. of Biomedical Engineering, Rensselaer Polytechnic Institute, New York, USA}
\affil[2]{Dept. of  Electrical and Computer Engineering, S\~{a}o Carlos School of Engineering, University of S\~{a}o Paulo, S\~{a}o Carlos, Brazil}
\affil[3]{Dept. of Medical Imaging, Hematology and Clinical Oncology, Ribeir\~{a}o Preto School of Medicine, University of S\~{a}o Paulo, Ribeir\~{a}o Preto, Brazil}

\affil[4]{Institute of Science and Technology for Brain-inspired Intelligence, Fudan University, Shanghai, China}

\affil[5]{Shanghai Center for Brain Science and Brain-inspired
Technology, Shanghai, China}

\affil[6]{Co-first authors}
\affil[7]{Co-corresponding authors}

\hypersetup{
pdftitle={},
pdfsubject={},
pdfauthor={},
pdfkeywords={},
}

\begin{document}

\twocolumn[ 
  \begin{@twocolumnfalse} 
  
\maketitle

\begin{abstract}
Digital mammography is still the most \textcolor{newaddcolor}{common imaging} tool for breast cancer screening. Although the benefits of using digital mammography for cancer screening outweigh the risks associated with the x-ray exposure, the radiation dose must be kept as low as possible while maintaining the diagnostic utility of the generated images, thus minimizing patient risks. Many studies investigated the feasibility of dose reduction by restoring low-dose images using deep neural networks. In these cases, choosing the appropriate training database and loss function is crucial and impacts the quality of the results. In this work, \textcolor{newaddcolor}{a modification of the ResNet architecture, with hierarchical skip connections, is proposed to restore low-dose digital mammography.} We compared \textcolor{newaddcolor}{the restored images to the standard full-dose images}. Moreover, we evaluated the performance of several loss functions for this task. For training purposes, \textcolor{newaddcolor}{we extracted 256,000 image patches from a dataset of 400 images of retrospective clinical mammography exams}, where different dose levels were simulated to generate low and standard-dose pairs. To validate the network in a real scenario, a physical anthropomorphic breast phantom was used to acquire real low-dose and standard full-dose images in \textcolor{newaddcolor}{a commercially avaliable mammography system}, which were then processed through our trained model. \textcolor{newaddcolor}{An analytical restoration model for low-dose digital mammography}, previously presented, was used as a benchmark in this work. \textcolor{newaddcolor}{Objective assessment was performed through the signal-to-noise ratio (SNR) and mean normalized squared error (MNSE), decomposed into residual noise and bias}. Results showed that the perceptual loss function (PL4) is able to achieve virtually the same noise levels of a full-dose acquisition, while resulting in smaller signal bias compared to other loss functions.  The source code for our deep neural network is available at \url{https://github.com/WANG-AXIS/LdDMDenoising}.
\end{abstract}


\vspace{0.35cm}

  \end{@twocolumnfalse} 
]

\section{Introduction}
\label{sec:introduction}

Early diagnosis of breast cancer is crucial to improve the survival rate. The expansion of breast screening programs contributed to improve this rate\textcolor{newaddcolor}{, which} has been significantly increased in the last years~\citep{saadatmand2015influence}. This disease is still the main cause of cancer-related deaths among women and \textcolor{newaddcolor}{screening mammography} is the primary tool for detecting tumors at early stages, especially for women at the age of 50-69~\citep{WHO2020}.

Full-field digital mammography (FFDM) and digital breast tomosynthesis (DBT) are the most common imaging tools for breast cancer screening~\citep{michell2018role}. In these systems a small dose of x-ray radiation is used to generate projections of the breast, which are then interpreted by a radiologist~\citep{vedantham2015digital}. Although the radiation dose levels are kept within a safe margin, it is desirable to keep the dose as low as possible while maintaining satisfied image quality to fulfill the clinical screening purposes~\citep{international2018iaea}. However, reducing radiation doses can degrade image quality, limiting the performance of the radiologist on searching and characterizing subtle lesions~\citep{haus2000screen,huda2003experimental, saunders2007does, chan2020effect}.

Several works in the field of medical imaging investigate the potential of reducing radiation dose by restoring low-dose (LD) exams to achieve image quality comparable to the ones at the clinical routine. Some proposals, in the field of computed tomography (CT)~\citep{kalra2003low, manduca2009projection, li2014adaptive} and digital mammography (DM)~\citep{wu2012dose, borges2017pipeline, borges2018restoration},  evaluated this technique in a model-based (MB) approach, using restoration methods through denoising techniques to improve image quality. In~\cite{borges2017pipeline} and~\cite{borges2018restoration}, the authors proposed a pipeline to restore LD mammography through a variance stabilizing transformation (VST). Recently, it was shown that the denoising method may improve the localization of microcalcifications (MC) in these exams~\citep{borges2020effect}.

With the rapid development of deep learning techniques, in particular the convolutional neural networks (CNNs), many studies have proposed algorithms to improve the quality of LD images and achieved comparable, or even better results, than the MB ones~\citep{wu2017iterative, kang2017deep, chen2017low1, wolterink2017generative, chen2017low2,  yang2018low, kang2018deep, shan20183, yang2018lowreports,  shan2019competitive, yin2019domain}. This new data-based approach takes advantage of learning features directly from a dataset rather than explicitly applying advanced feature extraction techniques or modeling the system mathematically. 

One important constraint of data-based techniques is the necessity of a great number of images \textcolor{newaddcolor}{and a diverse dataset} for the training process~\citep{sun2017revisiting}. In the field of medical imaging, access to large datasets is limited and some techniques, such as data augmentation and transfer learning, have been applied to increase the size of the dataset~\citep{lee2020deep, costa2020transfer}. Moreover, when it comes to LD image restoration using supervised learning, acquisitions at different dose levels are required in the training process. Although it may be possible to create experimental LD/FD imaging protocols, such as in~\cite{wolterink2017generative}, exposing the patient multiple times increases the risks of induced cancer.  

\textcolor{newaddcolor}{Thus, in the field of computed tomography (CT), a common approach is to train these deep networks using clinical data~\citep{mccollough2017low}, where LD images are obtained by injecting extra noise into the standard full-dose (FD) projections~\citep{shan2019competitive}. In the field of mammography, it is common to use breast specimen~\cite{liu2018radiation}, physical phantoms~\cite{gao2021deep} or virtual clinical trials (VCT) software~\citep{sahu2019using} to generate low-dose and full-dose image pairs for the training process of these deep networks.} \textcolor{reviewcolor}{In~\cite{green2019neural} the authors} adapted a noise injection technique from digital chest radiography to simulate ultra-low-dose mammography acquisitions and thus trained a CNN for denoising.

When it comes to the deep neural network (DNN) structure and the training step for the restoration process, there are two key components: the network architecture and loss function. The former determines the complexity of the denoising model, while the latter controls the learning process. \textcolor{newaddcolor}{Thus, the loss function has a direct impact on image quality and it is relatively more important than the network architecture for the task of image restoration~\citep{shan20183}}. 

As the restoration of LD mammography is composed of a denoising process, most image translation networks can be adapted to this task, such as residual encoder-decoder convolutional neural networks (RED-CNN)~\citep{chen2017low1}, U-net~\citep{ronneberger2015u} and dense networks~\citep{huang2017densely}. Even though the denoising process is part of the restoration process, the main goal of such a task is to map LD images to standard FD. That is where the loss function plays an important role in measuring the similarity between the image pair. Commonly-used loss functions include \textcolor{reviewcolor}{error visibility methods, such as} mean squared error (MSE) and mean absolute error (MAE); \textcolor{reviewcolor}{structural similarity methods, such as} structural similarity index (SSIM)~\citep{wang2004image} \textcolor{reviewcolor}{ and complex Wavelet SSIM index (CW-SSIM)~\citep{wang2005translation}; information-theoretical methods, such as IFC \citep{sheikh2005information} and VIF \cite{sheikh2006image}; and DNN methods, such as} perceptual loss (PL)~\citep{johnson2016perceptual}, adversarial loss~\citep{goodfellow2014generative} \textcolor{reviewcolor}{and image quality transformer (IQT)~\citep{cheon2021perceptual}}. \textcolor{reviewcolor}{In \citep{zhao2016loss}, the authors} investigated commonly-used losses for image restoration with neural networks for natural images. \textcolor{reviewcolor}{In \citep{ding2021comparison}, the authors also investigated several image quality assessment methods as loss function for low-level computer vision tasks on natural images}. 

However, in contrast with natural images, image restoration for medical images, more specifically for mammography, is a meticulous task, where subtle structures \textcolor{newaddcolor}{such as} MC are extremely important and must be preserved in the restoration process. Moreover, it is desired that the noise properties of the restored image match the FD ones, which is also important for radiologists in the clinical routine. \textcolor{reviewcolor}{This was demonstrated in~\citep{nagare2021bias}, where the authors proposed a loss function that takes into account the bias and noise variance for the restoration algorithm.}

The objective of this work is to investigate different loss functions of the DNN and their impact on the performance of the restoration of LD mammography images. \textcolor{newaddcolor}{We assess the efficiency of each loss function measuring the signal-to-noise ratio (SNR) and the mean normalized squared error (MNSE), decomposed into residual noise and bias, after the restoration process}. To this end, we propose a \textcolor{reviewcolor}{hierarchical residual convolutional neural network (HResNet)}. For the learning process, we build a \textcolor{newaddcolor}{clinical dataset from retrospective mammography exams, combined with a computational method to simulate LD acquisitions pairs}~\citep{borges2016method, borges2017method}. The adopted method performs noise injection in a variance-stabilizing domain, avoiding assumptions about the unknown noise-free signal. \textcolor{newaddcolor}{Furthermore, the spatial dependence of the quantum noise, the electronic noise and the noise spatial correlation were taken into account in the simulation, which enabled the generation of accurate clinical image samples for each patient, as if they had been acquired with lower radiation doses. At the end,} the assessment of the trained network was performed on \textcolor{newaddcolor}{real LD mammography acquisitions using a physical anthropomorphic breast phantom}.

\textcolor{reviewcolor}{The main contributions of this paper are summarized as follows.}
\begin{itemize}
	\item \textcolor{reviewcolor}{A comprehensive evaluation of common loss functions specifically to the field of medical imaging with an emphasis on DM.}
	\item \textcolor{reviewcolor}{The proposal for a new enhanced ResNet architecture, with hierarchical skip connections, extending the conventional one to better model the noise distribution of the LD images.}
	\item \textcolor{reviewcolor}{A new training strategy using 400 LD clinical data from retrospective mammography examinations was proposed. This enables the network to have contact with a diverse dataset, with different tissue structures, densities, signal and noise levels, etc.}
	\item \textcolor{reviewcolor}{Validation of each loss function with an objective metric that can evaluate blurring and noise individually.}
\end{itemize}

The \textcolor{newaddcolor}{remainder} of the paper is organized as follows. Section~\ref{sec:Theoretical_Background} introduces the theoretical background for the image degradation model and for the restoration model including both the MB and data-based, and the loss functions used in this study. Section~\ref{sec:Materials_Methods} presents the datasets used in this work, all the implementation details and the metrics used for the evaluation. In Section~\ref{sec:Results_Discussions}, the quantitative results and also \textcolor{newaddcolor}{some regions of interest (ROIs)} are shown for all the loss functions, followed by a concluding summary in Section~\ref{sec:Conclusion}.

\section{Theoretical Background}
\label{sec:Theoretical_Background}

The image restoration methods can be approached on several fronts. A common practice, which has been extensively presented in the literature, is to model the x-ray acquisition system and create mathematical formulations to perform the desired restoration. Although this was a common procedure in the past few years, new deep learning techniques have shown a great capacity of learning from the data, through supervised learning. This section presents the background information behind the image degradation model for mammography systems, the basics of the MB approach to restore the LD images and the proposed network. 

\subsection{Degradation Model}

Considering $\mat{X}\in\mathbb{R}^{w\times h} $ an observed X-ray mammography image of size $w\times h$ at standard FD, following~\citep{borges2018restoration}, we can model an acquisition as follows:
\begin{equation}
	\mat{X} = \mat{Y} + \mat{\eta}, \quad\mathrm{s.t.}\quad [\mat{\eta}]_{ij}\sim\mathcal{N}\left(0, \lambda\mat{Y}_{ij} + \sigma^2_e\right), 
	\label{eq:NoiseModel1}
\end{equation}
where $\mat{Y}$ is the noise-free image, and $\mat{\eta}$ is the corresponding noise at standard FD. Note that $[\mat{\eta}]_{ij}$ represents the element at the  $i$-th row and $j$-th column of the noise matrix $\mat{\eta}$, and follows a Gaussian distribution with zero mean and variance equal to $\lambda\mat{Y}_{ij} + \sigma^2_e$; here, $\lambda$ is the quantum noise gain and $\sigma^2_e$ is the variance of the electronic noise. Although in x-ray images the noise is often modeled through a Poisson-Gaussian distribution, the energy ranges at which DM operates allow the assumption of a signal-dependent Gaussian distribution, as done in Eq.~(\ref{eq:NoiseModel1}), thanks to the Central Limit Theorem~\citep{bertalmiodenoising2018}. Now let us consider another raw mammogram $\mat{X}_\gamma$, acquired using the same radiographic factors as $\mat{X}$ except for a reduction in current-time product (mAs), resulting in lower dose. The mammogram $\mat{X}_\gamma$ can be described as a function of the noise-free signal in Eq.~(\ref{eq:NoiseModel1}) as follows:
\begin{equation}
	\mat{X}_{\gamma} = \gamma \mat{Y} + \mat{\eta}_{\gamma},\quad\mathrm{s.t.}\quad [\mat{\eta}_{\gamma}]_{ij}\sim\mathcal{N}\left(0, \gamma\lambda\mat{Y}_{ij} + \sigma^2_e\right),
	\label{eq:NoiseModel2}
\end{equation}
where $0 < \gamma < 1$ is the mAs scaling factor, and thus the dose reduction factor.

The goal of this work is to investigate appropriate loss-functions capable of training a CNN to achieve Eq.~(\ref{eq:NoiseModel1}) starting from Eq.~(\ref{eq:NoiseModel2}), while keeping the noise-free signal $\mat{Y}$ as preserved as possible, with minimal blur and smear caused by the restoration process, \ie:
\begin{equation}
	\widehat{\mat{X}}=\Psi(\mat{X}_\gamma),
	\label{eq:NoiseModel3}
\end{equation}
where $\widehat{\mat{X}}$ is the restored image and $\Psi(\cdot)$ is the non-linear restoration operator.

\subsection{Restoration Model}

From Eqs.~(\ref{eq:NoiseModel1}) and~(\ref{eq:NoiseModel2}), the restoration task may be approached as a mathematical operator, which we will refer to as MB approaches. Alternatively, taking advantage of the great capacity of DNN, it is also possible to train a network to perform the whole restoration process avoiding the estimation of noise parameters. In this section we discuss both the model- and data-based approaches. 
\begin{figure*}[!tb]
	\centering
	\includegraphics[width=1\linewidth]{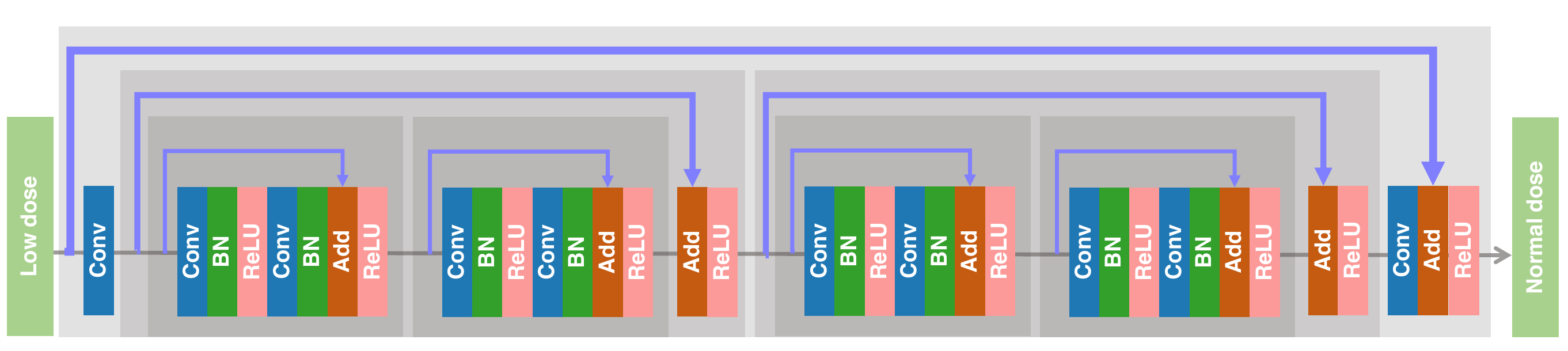}
	\caption{Architecture of the proposed network used in this study. }
	\label{fig:network}
\end{figure*}

\subsubsection{Model-based}

\textcolor{reviewcolor}{In \citep{borges2017pipeline,borges2018restoration}, our group proposed} a MB pipeline to restore LD mammography images, leveraging a variance stabilization technique, namely the generalized Anscombe transformation (GAT)~\citep{starck1998image}. With this technique, it is possible to use any denoising technique designed to treat signal-independent Gaussian-distributed data. The pipeline involves modeling the equipment's noise parameters, such as quantum noise gain, electronic noise variance, and the detector offset. These components are used in the GAT to bring the image to the domain where the noise model is signal-independent and approximately Gaussian. In the GAT domain, the block-matching and 3D filtering (BM3D)~\citep{dabov2007image} is used to suppress noise and the exact unbiased inverse of the generalized Anscombe transform is applied~\citep{makitalo2012optimal}. Finally, the denoised image is blended with the LD image through a weighted average. We refer to~\cite{borges2018restoration} for more details about the MB restoration process.

\subsubsection{Data-based}

This subsection presents the deep learning-based denoising model for restoration of LD DM. Despite the high dimensionality of the Euclidean space in medical images, it is known that these images lie in low-dimensional manifolds~\citep{wu2017iterative,yang2018low}. As DNNs, given sufficient trainable parameters, can approximate any non-linear transformation functions, one can model the non-linear transformation $\Psi$, from Eq.~(\ref{eq:NoiseModel3}), using deep neural-networks with the appropriate architecture and loss-function.

Inspired by the success of residual skip connection in various tasks such as image classification~\citep{he2016deep}, image denoising~\citep{wolterink2017generative,chen2017low1}, reinforcement learning~\citep{silver2017mastering}, we propose the HResNet, extending conventional ResNet to better model the noise distribution of the LD DM. Fig.~\ref{fig:network} shows the network architecture of the proposed restoration model used in this study.

In this architecture, the first level skip connection in the hierarchical structure connects the input and output of the entire network, which makes the network learn the noise field. In doing so, the denoising model has the following merits.
\begin{itemize}
	\item The output searching space is significantly reduced compared to the one without skip connection between the input and output. For example, when the images are normalized into the range of $[0,1]$, the conventional method without residual skip connection still has the same searching space as that of the ground-truth (GT), \ie, $[0,1]$. On the contrary,  our denoising model can decrease the output searching space by at least two orders of magnitude given the fact that noise is small relative to the attenuation.
	\item With such a skip connection, the network only needs to infer the residual between the LD and FD images, without memorizing the input itself.
	\item When the model is initialized by a Gaussian distribution with zero mean and a small variance, the network output before training is close to the input image, which can serve as a good staring point for optimization.
\end{itemize}

Another important aspect is that the network has one convolutional layer at the beginning to extract features from the input images, and one convolutional layer at the end to convert the learnt features into the noise field. In addition to that, the network has two second-level skip connections, each of them connects two residual blocks. The feature-maps from the first convolutional layer can be reused after two residual blocks. 

Last, the network has four residual blocks, each of them has a third-level skip connection and  serves as basic units. Each residual block contains four layers: two convolutional layers and two batch-normalization layers~\citep{ioffe2015batch}. The batch normalization layer has been proved to accelerate deep network training by reducing internal covariate shift. Rectified linear unit (ReLU) activation function is used after batch normalization layer or addition operation.  

We emphasize that the hierarchical skip connections from first level to third level represent the residual learning in a coarse to fine way. Note that the skip connection level can be further increased as one repeats this process; however, we used this three-level network as an example for restoration of LD DM in this study.

Specifically, all convolutional layers have 64 convolutional filters of size $3\times3$ with a stride of 1 and a zero-padding of 1 except for the final convolutional layer that has only one convolutional filter. Throughout the network, the feature-maps have the same size as the input image.

\subsection{Loss Functions}
\label{sec:LossFunctions}

The loss function plays an important role in training a restoration model and can determine the visual aspect of the generated images. \textcolor{reviewcolor}{In~\cite{zhao2016loss}, the authors} investigated the effects of several commonly-used losses for image restoration with neural networks. The difference between this study and theirs lies in two aspects. First, we study the effects of different losses for image restoration of LD DM, while~\cite{zhao2016loss} focused on natural images. In DM, as opposed to natural images, specific high-frequency and low-contrast features of the exams, such as MC and small masses, are vital for the successful clinical use of the data. Thus, in this scenario, it is preferable to retain some residual noise and keep those subtle features intact instead of aggressively filtering the data at the risk of masking lesions. Second, in addition to those metrics studied in~\cite{zhao2016loss}, we also studied the PL~\citep{johnson2016perceptual}, which is based on a pretrained VGG model~\citep{simonyan2014very}.

Different losses compute the similarity between the generated and GT images in different ways. We denote the generated and GT images as $\widehat{\mat{X}}\in\mathbb{R}^{w\times h}$ and $\mat{X}\in\mathbb{R}^{w\times h}$, respectively.  

\paragraph{{\bf Mean squared error  (MSE)}} MSE is the most widely used metric to measure the pixel-wise difference between generated and GT images. It can be formally defined as: 
\begin{equation}\label{eq:mse}
	\mathcal{L}_{\mathrm{MSE}} = \frac{1}{wh}\sum_{i=1}^w\sum_{j=1}^h \left([\widehat{\mat{X}}]_{ij} - [\mat{X}]_{ij}\right)^2,
\end{equation}
where $\mat{X}_{ij}$ indicates the element at $i$-th row and $j$-th column in $\mat{X}$.

\paragraph{{\bf Mean absolute error (MAE) or $\ell_1$}} Slightly different from MSE, MAE computes the $\ell_1$ loss between the generated and GT image. As a result, MAE does not over-penalize larger errors and can overcome the smoothness caused by MSE. It can be defined as:
\begin{equation}\label{eq:mae}
	\mathcal{L}_{\mathrm{MAE}} = \frac{1}{wh}\sum_{i=1}^w\sum_{j=1}^h \left|[\widehat{\mat{X}}]_{ij} - [\mat{X}]_{ij}\right|.
\end{equation}

\begin{figure*}[!htb]
	\centering	
	\subfloat[]{\includegraphics[scale=0.311]{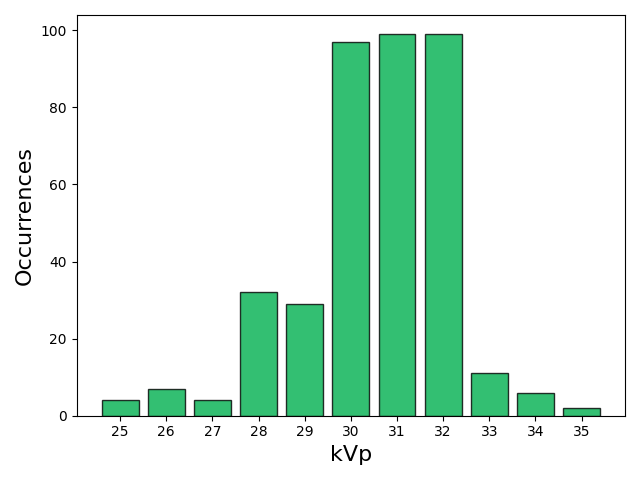}}
	\hfill
	\subfloat[]{\includegraphics[scale=0.311]{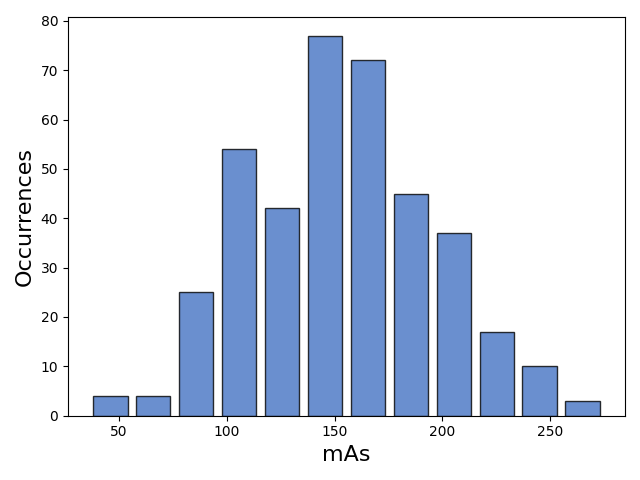}}
	\hfill
	\subfloat[]{\includegraphics[scale=0.311]{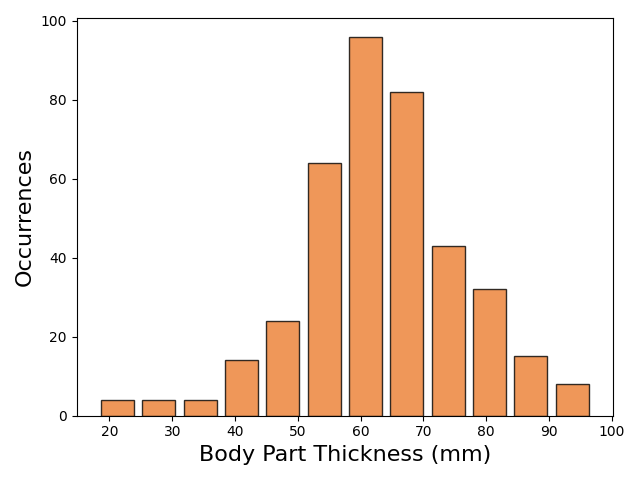}}
	\caption{\textcolor{reviewcolor}{Histograms showing the variability of (a) kVp, (b) mAs and (c) breast thickness in the clinical images from the training dataset.}}
	\label{fig:exames_statistics}
\end{figure*}

\paragraph{{\bf Structural similarity index (SSIM)}} SSIM is a widely-used image quality evaluation metric~\citep{wang2004image}. SSIM can measure the visual similarity between two images in terms of their structures and textures. The SSIM index is computed based on various windows of an image. The measure between the window $\widehat{\vct{x}}$ over $\widehat{\mat{X}}$ and the window $\vct{x}$ over $\mat{X}$, based on a common window size $k\times k$, can be defined as:  
\begin{equation}
	\mathrm{SSIM}(\widehat{\vct{x}},\vct{x})=\frac{2\mu_{\widehat{\vct{x}}}\mu_{\vct{x}}+c_1}{\mu_{\widehat{\vct{x}}}^2+\mu_{\vct{x}}^2+c_1}\frac{2\sigma_{\widehat{\vct{x}}\vct{x}}+c_2}{\sigma_{\widehat{\vct{x}}}^2+\sigma_{\vct{x}}^2+c_2},
\end{equation}
where $\mu_{\widehat{\vct{x}}}$ and $\mu_{\vct{x}}$ are the averages of $\widehat{\vct{x}}$ and $\vct{x}$ respectively, $\sigma_{\widehat{\vct{x}}}$ and $\sigma_{\vct{x}}$ are the variances of $\widehat{\vct{x}}$ and $\vct{x}$ respectively, and $\sigma_{\widehat{\vct{x}}\vct{x}}$ is the covariance of $\widehat{\vct{x}}$ and $\vct{x}$. Also, $c_1=1\times 10^{-4}$ and $c_2=9\times10^{-4}$ are two constants, which are used to stabilize the division with a weak denominator. The window size $k$ is 11, as suggested in~\cite{wang2004image}. The MSSIM between two images $\widehat{\mat{X}}$ and $\mat{X}$, $\mathrm{MSSIM}(\widehat{\mat{X}},\mat{X})$, refers to the average of the SSIM index over all windows. The SSIM loss is defined as: 
\begin{equation}
	\mathcal{L}_{\mathrm{SSIM}} = 1 - \mathrm{MSSIM}\left(\widehat{\mat{X}}, \mat{X}\right).
\end{equation}

\paragraph{{\bf Perceptual loss (PL)}} PL attempts to compare the similarity between two images in a high-level feature space~\citep{johnson2016perceptual}. A pretrained VGG model is widely used to extract features from an image to form such a high-level feature space, which is expected mimic the human visual system. PL is similar to MSE, but in a feature space instead of pixel space. It can be defined as: 
\begin{equation}\label{eq:pl}
	\mathcal{L}_{\mathrm{PL}} = \frac{1}{w'h'c'}\sum_{i=1}^{w'}\sum_{j=1}^{h'}\sum_{k=1}^{c'} \left([\mat{\Phi}(\widehat{\mat{X}})]_{ijk} - [\mat{\Phi}(\mat{X})]_{ijk}\right)^2,
\end{equation}
where $\mat{\Phi}$ represents the feature extractor, whose output is a tensor of size $w'\times h'\times c'$. \textcolor{reviewcolor}{The PL can be computed on early or later layers of the VGG network. Each layer is commonly denominated as a block of convolutions and activation functions, \eg ReLU, before the max-pooling. For example, PL1 contains the first two convolution layers and their respective activation function.}

\section{Materials \& Methods}
\label{sec:Materials_Methods}

For reliable training and validation, we created two distinct sets of images: the first with clinical cases and the second with anthropomorphic breast phantom images.  We used a hybrid dataset of clinical images with simulated LD/FD pairs and later restricted testing the trained network in the phantom data, with LD acquisitions acquired directly in the mammography equipment. Doing so, we avoid the so-called “inverse crime”, which is to test and train the model with the same fabricated synthetic data~\citep{zhao2016loss}. In this section, we specify how both datasets used in this study were constructed.

Also, in this section we present the DNN implementation details and the evaluation metrics used in this study.

\subsection{Training Dataset}

The training dataset consists of 400 clinical mammography acquired at the Barretos Cancer Hospital (Brazil). \textcolor{newaddcolor}{These data were obtained retrospectively from breast cancer screening examinations, after approval by the institutional review board}. This dataset is relative to 100 patients with their respective images from the craniocaudal (CC) and mediolateral oblique (MLO) views of the left and right breast. \textcolor{newaddcolor}{All images was acquired using a} Hologic Selenia Dimensions Mammography System (Hologic, Bedford, MA) and all the images were saved as raw data, \ie, DICOM ``\textit{for processing}''. \textcolor{reviewcolor}{Fig.~\ref{fig:exames_statistics} shows the occurrence of different values of kVp, mAs, and breast thickness in the training dataset.} All clinical images were fully anonymized to preserve patients' medical records.

As previously discussed, exposing patients to x-ray radiation several times to build an image dataset with LD acquisitions can be dangerous and impractical in the clinical routine. In order to acquire clinical images at lower doses, we used a previous work to simulate dose reduction in these data~\citep{borges2016method, borges2017method}. The method injects quantum and electronic noise in the VST domain and also accounts for the detector crosstalk. We refer to both these works for further details about this technique. We applied this technique to all clinical images to simulate acquisitions of 75\% and 50\% of the standard FD, in which the image was originally obtained ($\gamma = 0.75$ and $\gamma = 0.50$, respectively). After the simulation, we reached a total of 1,200 images among full and reduced doses.

\subsection{Testing Dataset}

\begin{figure}[!t]
	\begin{center}
		\includegraphics[width=1.0\linewidth]{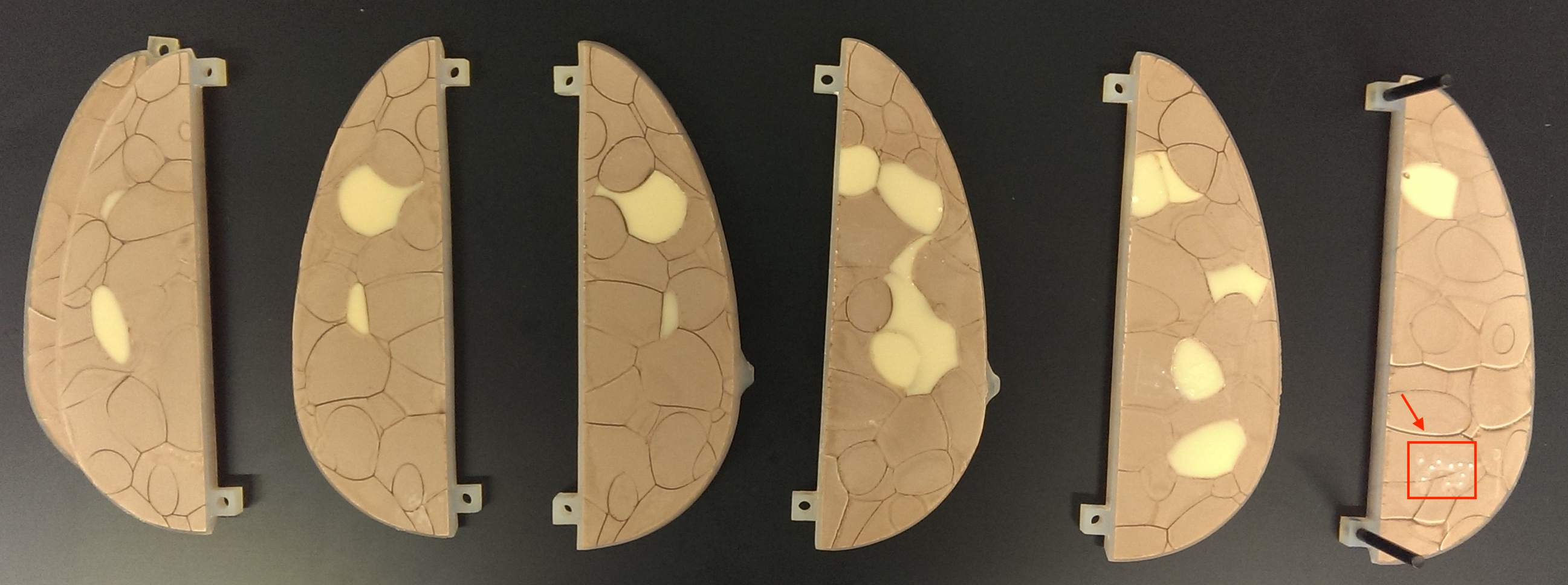}
	\end{center}
	\caption{Image of the physical anthropomorphic breast phantom used in this study. The red arrow points to the pieces of calcium oxalate that were placed between the slabs to simulate a cluster of microcalcifications.}
	\label{fig:Anthropomorphic_phantom}
\end{figure}

To validate the restoration methods that were trained on the clinical dataset, we acquired images of a physical anthropomorphic breast phantom at the Hospital of the University of Pennsylvania. \textcolor{reviewcolor}{We also used a Hologic Selenia Dimensions Mammography System, as for the training dataset. The phantom has six slabs, with total thickness of 51 mm.} It consists of a material that mimics the breast tissue and was prototyped by CIRS, Inc. (Reston, VA), under the license from the University of Pennsylvania~\citep{carton2011development}. Small pieces of calcium oxalate (99\%, Alfa Aesar, Ward Hill, MA) were placed between the six slabs to simulate MCs, as illustrated in Fig.~\ref{fig:Anthropomorphic_phantom}.

\textcolor{newaddcolor}{We acquired a total of 25 images} of the anthropomorphic breast phantom. Firstly, for the FD, the automatic exposure control (AEC) was used to select the standard radiographic factors, which yielded a combination of 29 kVp and 160 mAs. Without physically moving the phantom, the system was set to manual mode and 15 exposures at the standard radiographic factors were performed to generate a set of FD images. For reduction factors of 75\% and 50\%, \textcolor{newaddcolor}{10 images were acquired by reducing the current-time product from 160~mAs to 120~mAs and 80~mAs respectively}. It is important to note that all these images were saved as raw data.

\subsection{Figures of Merit}
\label{cap:FOM}
To take advantage of being capable of exposing the physical anthropomorphic breast phantom several times\textcolor{newaddcolor}{, we generated} a pseudo-GT and compared all the \textcolor{newaddcolor}{restored images} with this GT through an assessment of the signal-to-noise ratio (SNR) and the mean normalized squared error (MNSE) decomposed into residual noise ($\mathcal{R}_{\mathcal{N}}$) and bias squared ($\mathcal{B}^2$). This metric was previously presented in~\cite{borges2018restoration} \textcolor{newaddcolor}{for the assessment of digital breast tomosynthesis images}.

\textcolor{reviewcolor}{We first measured the signal-to-noise ratio (SNR) in all phantom images and its restorations as the ratio of the mean pixel value and its standard deviation along the five realizations. The metric was calculated only inside the breast region and an average filter of size 15$\times$15 was used to smooth both the mean signal value and the standard deviation after their calculation.}

Considering $\mathcal{X} \subset\mathbb{R}^{w\times h} $ a subspace of FD mammography acquisitions, suppose we have a set of $N$ realizations $\mat{X}^{*} \in \mathcal{X}$ for the the GT estimation, and a set of $P$ realizations $\mat{X}^{'} \in \mathcal{X}$ for MNSE assessment. Here we refer to the GT as the expectation of the acquisitions on this subspace $\widehat{\mat{X}}_{GT} = \mathbb{E}\{\mat{X}^{*}\}$. Also, after breast phantom segmentation, we denote $(i,j)$ as a pair of 2D indices running inside a set $\mathcal{I} \in\mathbb{X}^{+}$ of size $\mat{I} < w \times h$. The set $\mathcal{I}$ represents the collection of pixel coordinates after the segmentation. The MNSE can be calculated as:
\begin{align}
	\mathrm{MNSE} =  &\frac{1}{P}\sum_{p=1}^{P}
	\underbrace{\frac{1}{\mat{I}}\sum_{(i,j)\,\in\, \mathcal{I}}{\frac{\left([\mat{X}^{'}_{p}]_{ij} - 
				[\widehat{\mat{X}}_{GT}]_{ij}\right)^2}{[\widehat{\mat{X}}_{GT}]_{ij}}}}_{\mathrm{NQE}} \notag\\ &- \underbrace{\frac{1}{\mat{I}N} \sum_{(i,j)\,\in\,\mathcal{I}}{\frac{\mathbb{V} \left([\mat{X}^{*}]_{ij}\right)}{[\widehat{\mat{X}}_{GT}]_{ij}}}}_{\phi_1},
\end{align}
where $\phi_{1}$ is accounting for the error associated with the limited number of images used for the GT estimation and $\mathbb{V} \left(\mat{X}^{*}\right)$ is a point-wise variance among this set of images. It is possible to decompose the MNSE into $\mathcal{R}_{\mathcal{N}}$ and $\mathcal{B}^2$ portions, such that:
\begin{align}
	\mathrm{MNSE} = &\underbrace{\left\{ \frac{1}{\mat{I}}\sum_{(i,j)\,\in\,\mathcal{I}}{\frac{\left(\mathbb{E}\{[\mat{X}^{'}]_{ij}\} - [\widehat{\mat{X}}_{GT}]_{ij}\right)^2}{[\widehat{\mat{X}}_{GT}]_{ij}}} \right\} - \phi_{1} - \frac{\mathcal{R}_{\mathcal{N}}}{P}}_{\mathcal{B}^2} \notag\\ & + \underbrace{\frac{1}{\mat{I}}\sum_{(i,j)\,\in\,\mathcal{I}}{\frac{\mathbb{V}\left([\mat{X}^{'}]_{ij}\right)}{[\widehat{\mat{X}}_{GT}]_{ij}}}}_{\mathcal{R}_{\mathcal{N}}}.
\end{align}

\begin{figure}[!t]
	\begin{center}
		\includegraphics[width=1.0\linewidth]{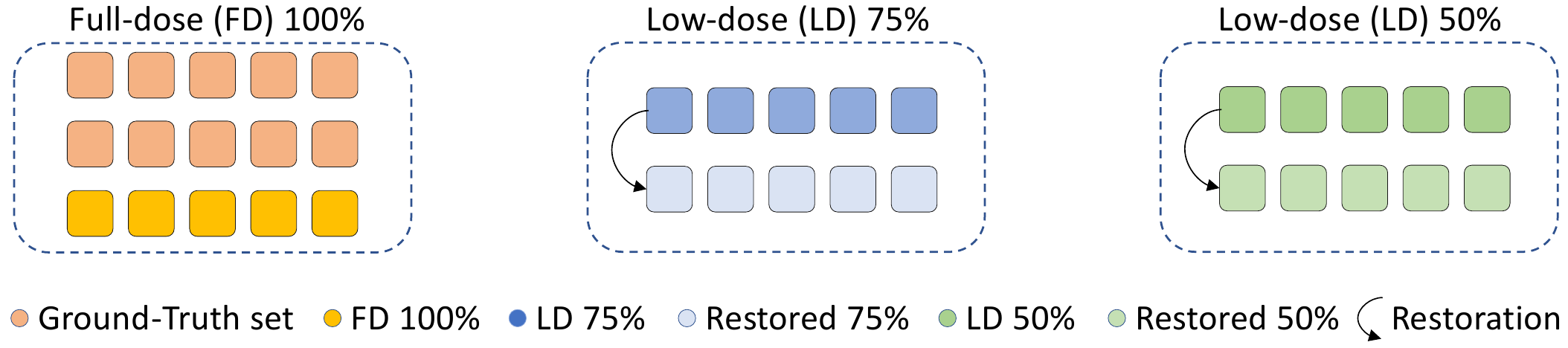}
	\end{center}
	\caption{Illustration of how the anthropomorphic breast phantom dataset was subdivided for the assessment of the MNSE and its decomposition. The metrics were evaluated on all different colored sub-groups of FD, LD and restored images.}
	\label{fig:imgScheme_MNSE}
\end{figure} 
For the practical evaluation, we first manually segmented the phantom image to avoid calculating the metric outside the breast tissue. From the 15 images at the FD, 10 were used to generate the GT ($N = 10$) and 5 to calculate the metric ($P = 5$). As previously mentioned, 5 images of each reduced dose ($P_{\gamma} = 5$) were used to calculate the metric as well. The scheme shown in Fig.~\ref{fig:imgScheme_MNSE} illustrates how we separated all the phantom images. The MNSE and its decomposition were evaluated on all different colored sub-groups of LD and restored images.

To avoid error on the $\mathcal{B}^2$ due to differences in the mean value from different acquisitions, we used a technique of fitting a first-order polynomial to correct the image mean value. First, the GT was calculated following the previous equations, and then all the images used to generate the GT itself were adjusted based on the calculated GT, as done in~\cite{borges2018restoration}. After this correction, the GT was calculated again and all the non-GT FD acquisitions were also adjusted through the fitting technique.  All the restored images and the LD acquisitions had their mean value adjusted through this method. This overall process guarantees that the $\mathcal{B}^2$ error measurement is due to the smearing/blurring imposed by the restoration process and not small changes in the mean value of the images.

We also compared the time spent to run the restoration process on the model based approach and in the deep network.

\subsection{Implementation Details}

Overall, we trained 14 networks in this study---7 dedicated to restore 75\%-dose images and 7 dedicated to 50\%-dose. Each of these 7 networks refer to the mentioned loss functions from Section~\ref{sec:LossFunctions}. \textcolor{reviewcolor}{Although there is a range of different loss functions for image restoration, as shown in \citep{ding2021comparison}, we used in this work examples of error visibility methods, \eg, the MAE, structural similarity method, \eg, the SSIM, and DNN method, \eg, the PL. These losses are commonly used by previous work in the field of medical imaging restoration \citep{chen2017low1, chen2017low2, yang2018low, kang2018deep, shan20183, yang2018lowreports, shan2019competitive}}. For the PL, following~\cite{johnson2016perceptual}, we used VGG-16 network~\citep{simonyan2014very}, which was pretrained on the ImageNet dataset and publicly available from PyTorch official website. Since different layers of the VGG-16 networks can form different feature space, we used the feature-maps right before the first four max-pooling layers to form four different feature spaces, and then obtain four corresponding PLs from shallow to deep. We denote these four PLs as $\mathcal{L}_{\mathrm{PL1}}$, $\mathcal{L}_{\mathrm{PL2}}$, $\mathcal{L}_{\mathrm{PL3}}$, and $\mathcal{L}_{\mathrm{PL4}}$. We found that too many max-pooling layers involved in PL largely affect pixel-wise comparison, therefore, we removed all max-pooling layers in PL4.

For all neural networks with different losses, the initial learning rate $\lambda$ was set as $1.0\times10^{-4}$ and was reduced half by every $10$ epochs. The trainable parameters in the network was optimized using the Adam optimization~\citep{kingma2014adam}, whose coefficients used for computing running averages of gradients and its square were set as $0.5$ and $0.999$, respectively. The network was implemented with PyTorch DL library~\citep{paszke2017automatic} and trained within $60$ epochs using a NVIDIA GeForce GTX 1080 Ti GPU. The batch size was set as 256 during the training; however, it was reduced accordingly for the training with PL since VGG network also needs to be in the GPU. \textcolor{newaddcolor}{As PLs-based networks are hard to train, we used a pre-trained MAE network as the starting point.}
\begin{figure}[!t]
	\begin{center}
		\includegraphics[width=1.0\linewidth]{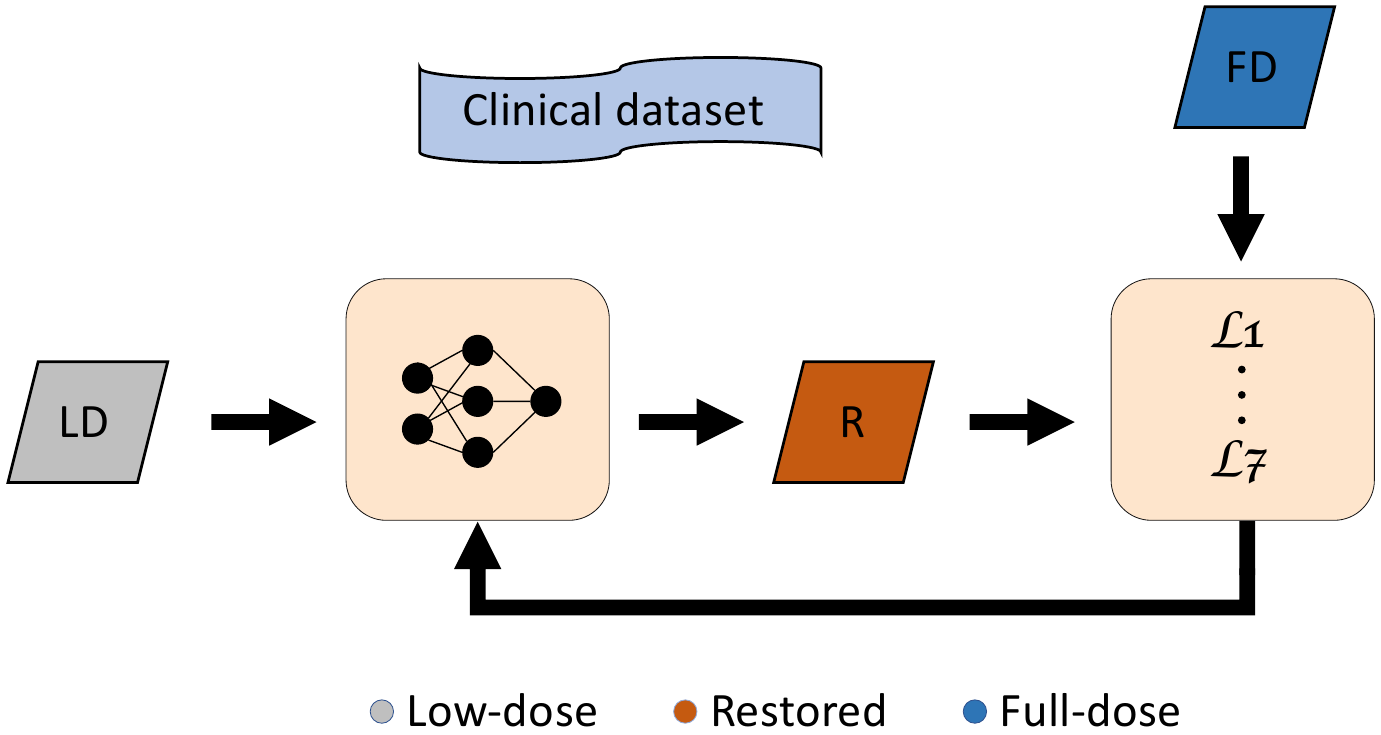}
	\end{center}
	\caption{\textcolor{newaddcolor}{Schematic explaining the training process of the network}.}
	\label{fig:imgScheme_Train}
\end{figure} 

\begin{figure}[t]
	\begin{center}
		\includegraphics[width=1.0\linewidth]{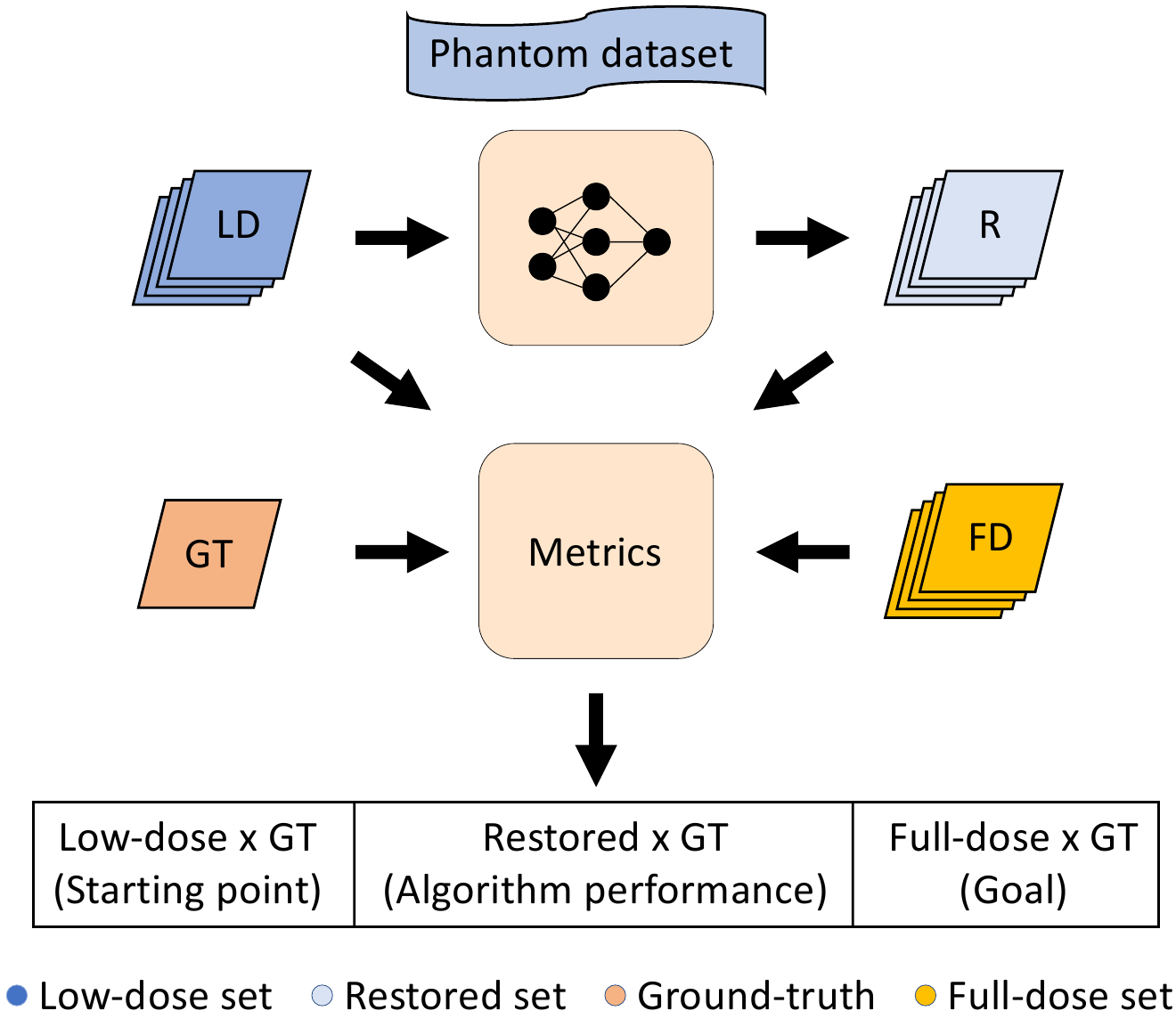}
	\end{center}
	\caption{\textcolor{newaddcolor}{Schematic explaining the test procedure for the network.}}
	\label{fig:imgScheme_Test}
\end{figure} 

To train the restoration model, a total of 256,000 patches of size 64$\times$64 were randomly selected from the breast regions of the 400 clinical images available in this study. These patches include pairs of LD and FD images, \ie, we trained one neural network for each reduction factor.

After the training process, \textcolor{newaddcolor}{which is illustrated in Fig.~\ref{fig:imgScheme_Train},} the network was quantitatively tested on the anthropomorphic breast phantom dataset to validate the effects of different loss functions. \textcolor{reviewcolor}{The testing stage, illustrated in Fig.~\ref{fig:imgScheme_Test}, involved measuring the MNSE between the GT against the LD, FD and restored images. The value from the LD/GT gives us the starting point of $\mathcal{R}_{\mathcal{N}}$ and $\mathcal{B}^2$. At this stage, $\mathcal{R}_{\mathcal{N}}$ is usually high and $\mathcal{B}^2$ low. The FD/GT guide us to the goal of the restoration regarding the lowest $\mathcal{B}^2$ and the desired $\mathcal{R}_{\mathcal{N}}$ level. When evaluating the Restored/GT, it is possible to see how the restoration performed, \ie, how much blurring was imposed, measuring through the $\mathcal{B}^2$, and how close the $\mathcal{R}_{\mathcal{N}}$ is to the FD value. Implementation details of how this metric was calculated on each of these groups are presented in Section \ref{cap:FOM}.}

\section{Results \& Discussions}
\label{sec:Results_Discussions}

\textcolor{reviewcolor}{In this section, we present and discuss the quantitative assessment performed on the phantom images and also the ROIs for both clinical and phantom images. We compared the LD and FD acquisitions to the images restored by the proposed network using the folowing loss functions: MSE, MAE, SSIM, PL1, PL2, PL3 and PL4. Also, for comparison with an analytical method previously published in the literature, we used the MB approach proposed in \citep{borges2018restoration} as a benchmark for all different loss functions of the proposed DNN.}

\subsection{Visual \textcolor{reviewcolor}{Analysis}}

\begin{figure*}[!htb]
	\centering	
	\subfloat[FD]{\includegraphics[scale=2.1]{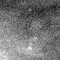}}
	\hfill
	\subfloat[LD]{\includegraphics[scale=2.1]{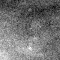}}
	\hfill
	\subfloat[MB]{\includegraphics[scale=2.1]{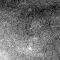}}
	\hfill
	\subfloat[$\mathcal{L}_{\mathrm{MSE}}$]{\includegraphics[scale=2.1]{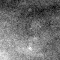}}
	\hfill
	\subfloat[$\mathcal{L}_{\mathrm{MAE}}$]{\includegraphics[scale=2.1]{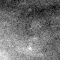}}
	\hfill
	
	\subfloat[$\mathcal{L}_{\mathrm{SSIM}}$]{\includegraphics[scale=2.1]{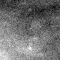}}
	\hfill
	\subfloat[$\mathcal{L}_{\mathrm{PL1}}$]{\includegraphics[scale=2.1]{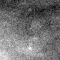}}
	\hfill
	\subfloat[$\mathcal{L}_{\mathrm{PL2}}$]{\includegraphics[scale=2.1]{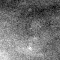}}
	\hfill
	\subfloat[$\mathcal{L}_{\mathrm{PL3}}$]{\includegraphics[scale=2.1]{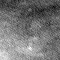}}
	\hfill
	\subfloat[$\mathcal{L}_{\mathrm{PL4}}$]{\includegraphics[scale=2.1]{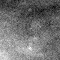}}
	
	\caption{Illustration of a magnified ROI of a clinical image focusing a small cluster of MC. (a) FD acquisition; (b) simulated LD acquisition for a dose reduction factor of 75\%; restored images generated by the (c) MB method; (d) proposed network with the loss function MSE, (e) MAE, (f) SSIM and (g)-(j) PL1 to PL4, respectively. All the images were normalized based on the FD and are displayed in the same dynamic range.}
	\label{fig:clinical_75_imgCap6BR3D}
\end{figure*}

\begin{figure*}[!htb]
	\centering	
	\subfloat[FD]{\includegraphics[scale=2.1]{imgs/clinical_ROIs/838542_Mammo_L_CC.png}}
	\hfill
	\subfloat[LD]{\includegraphics[scale=2.1]{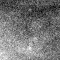}}
	\hfill
	\subfloat[MB]{\includegraphics[scale=2.1]{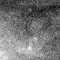}}
	\hfill
	\subfloat[$\mathcal{L}_{\mathrm{MSE}}$]{\includegraphics[scale=2.1]{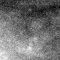}}
	\hfill
	\subfloat[$\mathcal{L}_{\mathrm{MAE}}$]{\includegraphics[scale=2.1]{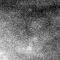}}
	
	\subfloat[$\mathcal{L}_{\mathrm{SSIM}}$]{\includegraphics[scale=2.1]{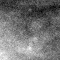}}
	\hfill
	\subfloat[$\mathcal{L}_{\mathrm{PL1}}$]{\includegraphics[scale=2.1]{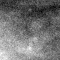}}
	\hfill
	\subfloat[$\mathcal{L}_{\mathrm{PL2}}$]{\includegraphics[scale=2.1]{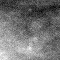}}
	\hfill
	\subfloat[$\mathcal{L}_{\mathrm{PL3}}$]{\includegraphics[scale=2.1]{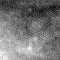}}
	\hfill
	\subfloat[$\mathcal{L}_{\mathrm{PL4}}$]{\includegraphics[scale=2.1]{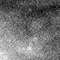}}
	
	\caption{Illustration of a magnified ROI of a clinical image focusing a small cluster of MC. (a) FD acquisition; (b) simulated LD acquisition for a dose reduction factor of 50\%; restored images generated by the (c) MB method; (d) proposed network with the loss function MSE, (e) MAE, (f) SSIM and (g)-(j) PL1 to PL4, respectively. All the images were normalized based on the FD and are displayed in the same dynamic range.}
	\label{fig:clinical_50_imgCap6BR3D}
\end{figure*}

\begin{figure*}[!htb]
	\centering	
	\subfloat[FD]{\includegraphics[scale=1.58]{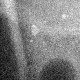}}
	\hfill
	\subfloat[LD]{\includegraphics[scale=1.58]{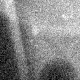}}
	\hfill
	\subfloat[MB]{\includegraphics[scale=1.58]{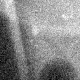}}
	\hfill
	\subfloat[$\mathcal{L}_{\mathrm{MSE}}$]{\includegraphics[scale=1.58]{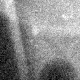}}
	\hfill
	\subfloat[$\mathcal{L}_{\mathrm{MAE}}$]{\includegraphics[scale=1.58]{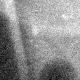}}
	
	\subfloat[$\mathcal{L}_{\mathrm{SSIM}}$]{\includegraphics[scale=1.58]{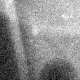}}
	\hfill
	\subfloat[$\mathcal{L}_{\mathrm{PL1}}$]{\includegraphics[scale=1.58]{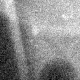}}
	\hfill
	\subfloat[$\mathcal{L}_{\mathrm{PL2}}$]{\includegraphics[scale=1.58]{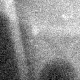}}
	\hfill
	\subfloat[$\mathcal{L}_{\mathrm{PL3}}$]{\includegraphics[scale=1.58]{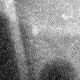}}
	\hfill
	\subfloat[$\mathcal{L}_{\mathrm{PL4}}$]{\includegraphics[scale=1.58]{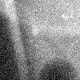}}
	
	\caption{Illustration of a magnified ROI of an anthropomorphic breast phantom image focusing a simulated cluster of MC. (a) FD acquisition; (b) LD acquisition for a dose reduction factor of 75\%; restored images generated by the (c) MB method; (d) proposed network with the loss function MSE, (e) MAE, (f) SSIM and (g)-(j) PL1 to PL4, respectively. All the images were normalized based on the FD and are displayed in the same dynamic range.}
	\label{fig:phantom_75_imgCap6BR3D}
\end{figure*}

\begin{figure*}[!htb]
	\centering	
	\subfloat[FD]{\includegraphics[scale=1.58]{imgs/phantom_ROIs/160_01_Mammo_R_CC.png}}
	\hfill
	\subfloat[LD]{\includegraphics[scale=1.58]{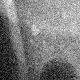}}
	\hfill
	\subfloat[MB]{\includegraphics[scale=1.58]{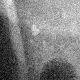}}
	\hfill
	\subfloat[$\mathcal{L}_{\mathrm{MSE}}$]{\includegraphics[scale=1.58]{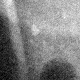}}
	\hfill
	\subfloat[$\mathcal{L}_{\mathrm{MAE}}$]{\includegraphics[scale=1.58]{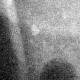}}
	
	\subfloat[$\mathcal{L}_{\mathrm{SSIM}}$]{\includegraphics[scale=1.58]{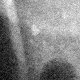}}
	\hfill
	\subfloat[$\mathcal{L}_{\mathrm{PL1}}$]{\includegraphics[scale=1.58]{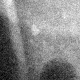}}
	\hfill
	\subfloat[$\mathcal{L}_{\mathrm{PL2}}$]{\includegraphics[scale=1.58]{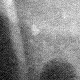}}
	\hfill
	\subfloat[$\mathcal{L}_{\mathrm{PL3}}$]{\includegraphics[scale=1.58]{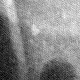}}
	\hfill
	\subfloat[$\mathcal{L}_{\mathrm{PL4}}$]{\includegraphics[scale=1.58]{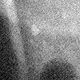}}
	\hfill
	
	\caption{Illustration of a magnified ROI of an anthropomorphic breast phantom image focusing a simulated cluster of MC. (a) FD acquisition; (b) LD acquisition for a dose reduction factor of 50\%; restored images generated by the (c) MB method; (d) proposed network with the loss function MSE, (e) MAE, (f) SSIM and (g)-(j) PL1 to PL4, respectively. All the images were normalized based on the FD and are displayed in the same dynamic range.}
	\label{fig:phantom_50_imgCap6BR3D}
\end{figure*}

Figs.~\ref{fig:clinical_75_imgCap6BR3D} and~\ref{fig:clinical_50_imgCap6BR3D} display a magnified ROI of a clinical image acquired at the standard radiation dose and at the dose reduction factors of 75\% and 50\%, respectively. As expected, the LD mammography presents more perceived noise as compared with the FD mammography. The restored results of LD mammography are shown in (c)-(j). We chose an ROI with a MC cluster, which is an important feature for cancer diagnosis. As expected, for the reduction factor of 50\% (Fig.~\ref{fig:clinical_50_imgCap6BR3D}), we can see that the loss functions MSE and MAE yield an overall smoothing in the image. For the SSIM, there is a subtle difference in the noise level when compared with MAE and MSE. Again, when we go from PL1 to PL4, it is visually noticeable that the blurring effect decreases and the fine details are preserved. Moreover, we can recognize some texture artifacts in the PL3 image, which can explain the higher $\mathcal{B}^2$ value, contrasting with the tendency to decrease this value coming from PL1 to PL4. It is important to note the visual similarity between the DNN with PL4 loss function and the MB method. All the visual \textcolor{reviewcolor}{analysis} discussed agree with the quantitative results presented in the next section. The same \textcolor{newaddcolor}{discussion} can be can be done for the 75\% case, in Fig.~\ref{fig:clinical_75_imgCap6BR3D}; however, the differences are more subtle when compared with the 50\%.

\begin{figure*}[htb]
	\centering	
	\subfloat[FD]{\includegraphics[scale=0.115]{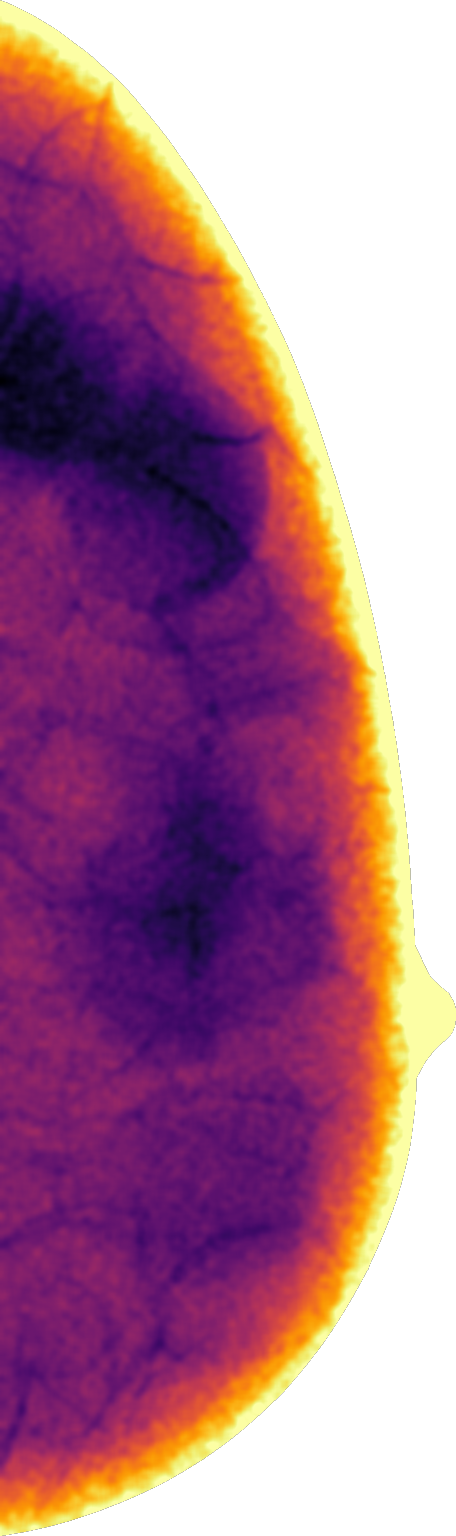}}
	\subfloat[LD]{\includegraphics[scale=0.115]{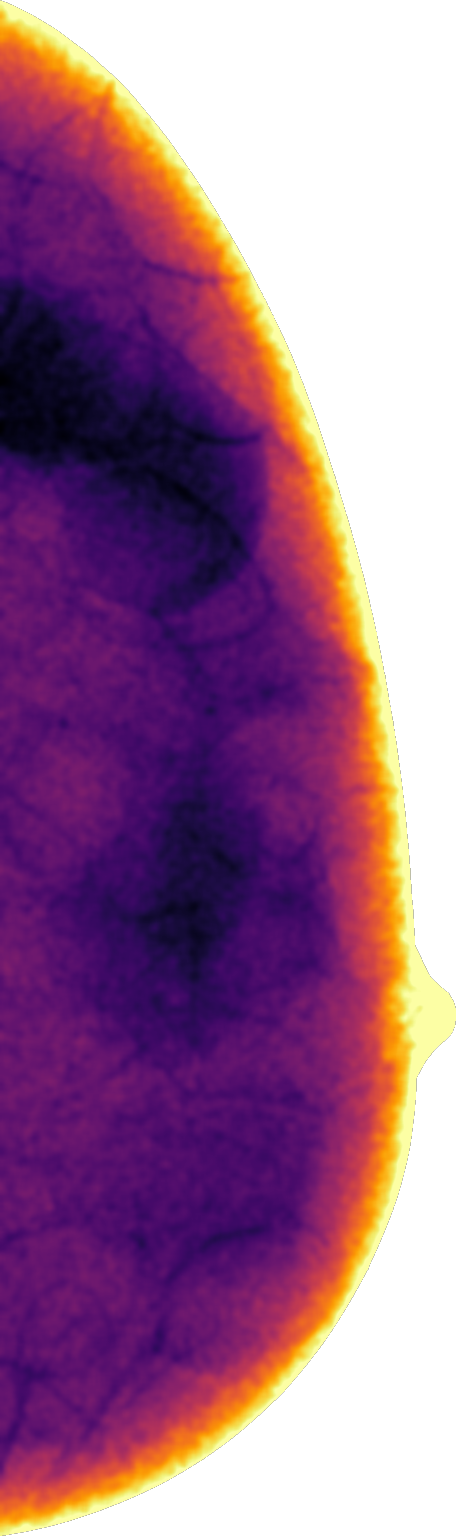}}
	\subfloat[MB]{\includegraphics[scale=0.115]{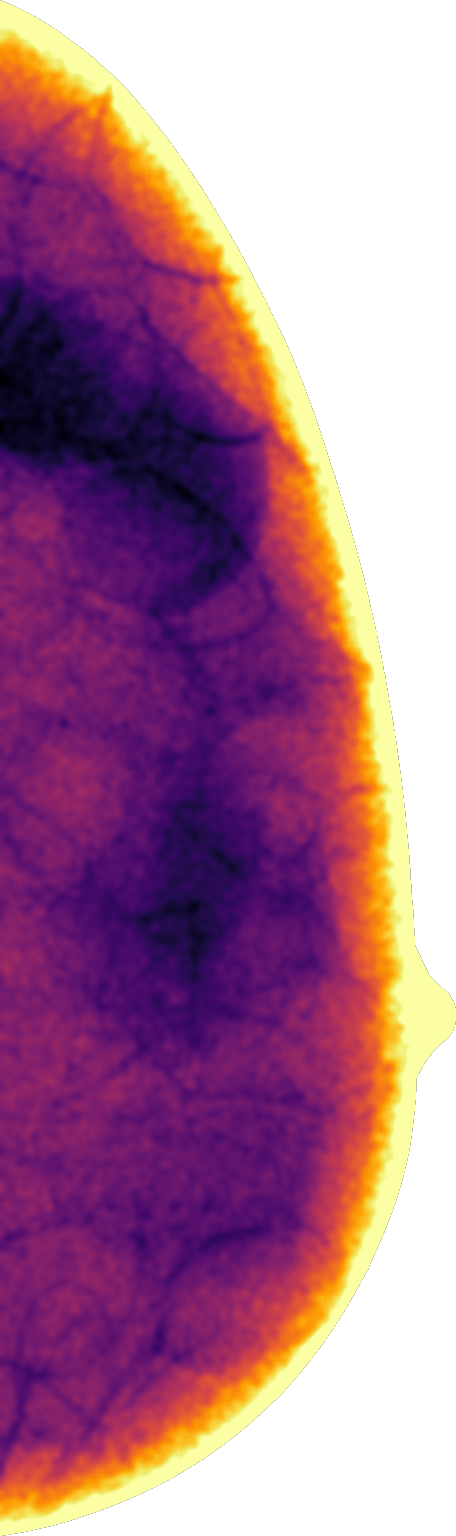}}
	\subfloat[$\mathcal{L}_{\mathrm{MSE}}$]{\includegraphics[scale=0.115]{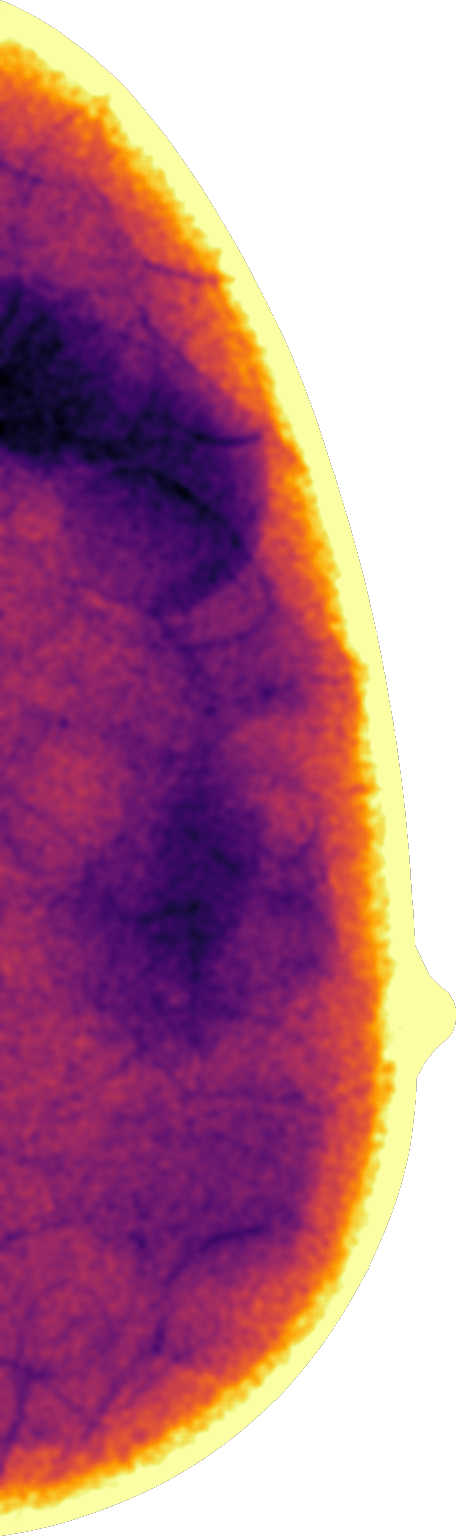}}
	\subfloat[$\mathcal{L}_{\mathrm{MAE}}$]{\includegraphics[scale=0.115]{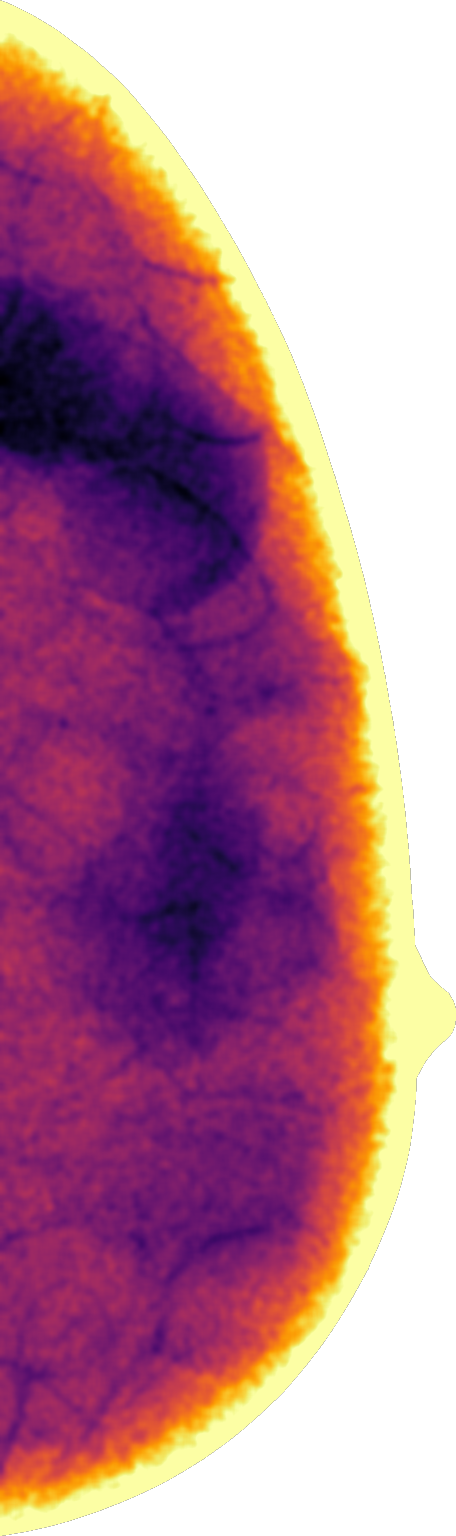}}
	\subfloat[$\mathcal{L}_{\mathrm{SSIM}}$]{\includegraphics[scale=0.115]{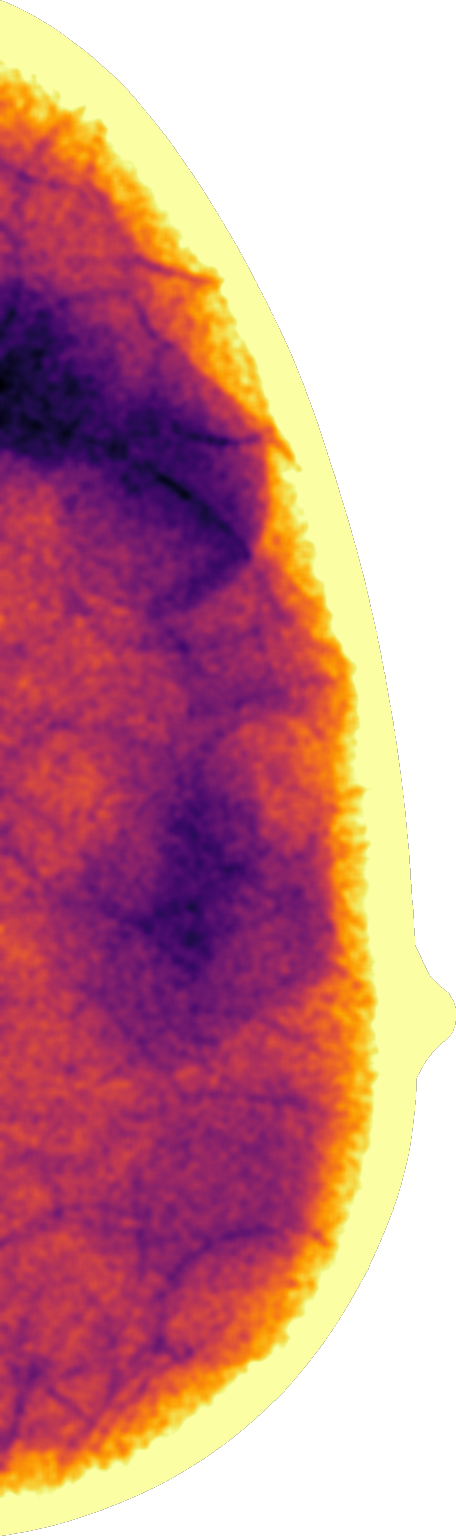}}
	\subfloat[$\mathcal{L}_{\mathrm{PL1}}$]{\includegraphics[scale=0.115]{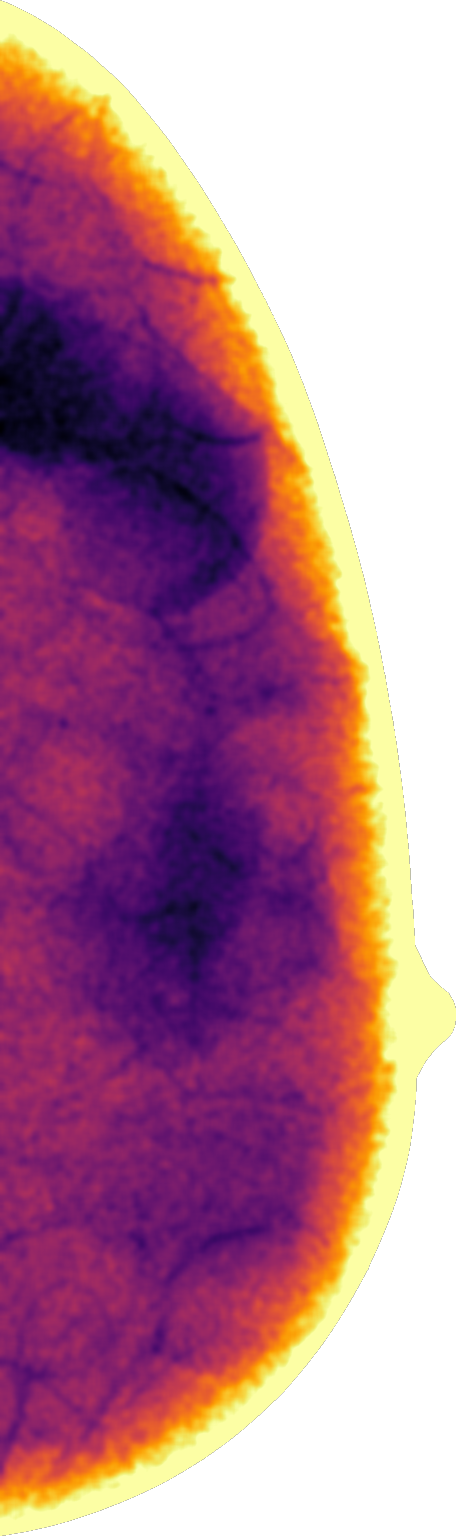}}
	\subfloat[$\mathcal{L}_{\mathrm{PL2}}$]{\includegraphics[scale=0.115]{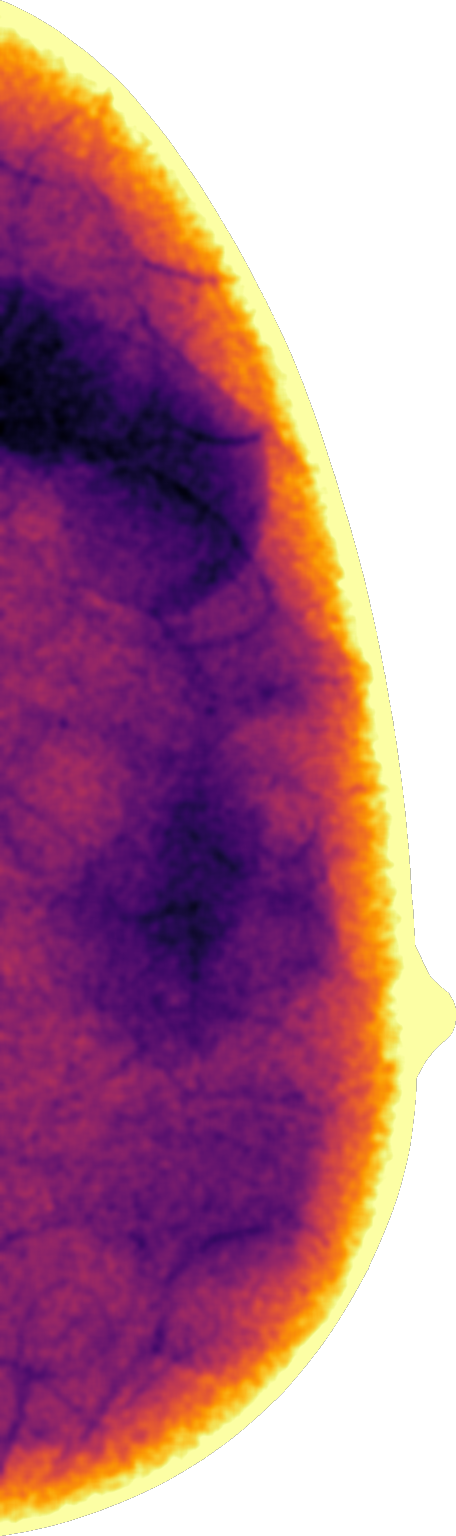}}
	\subfloat[$\mathcal{L}_{\mathrm{PL3}}$]{\includegraphics[scale=0.115]{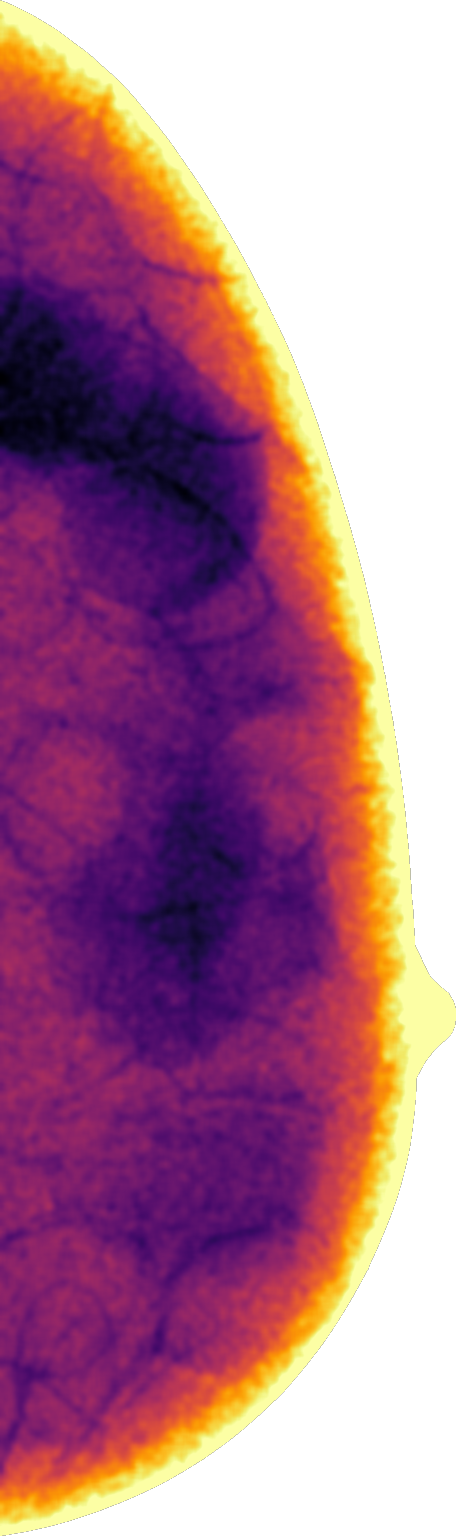}}
	\subfloat[$\mathcal{L}_{\mathrm{PL4}}$]{\includegraphics[scale=0.115]{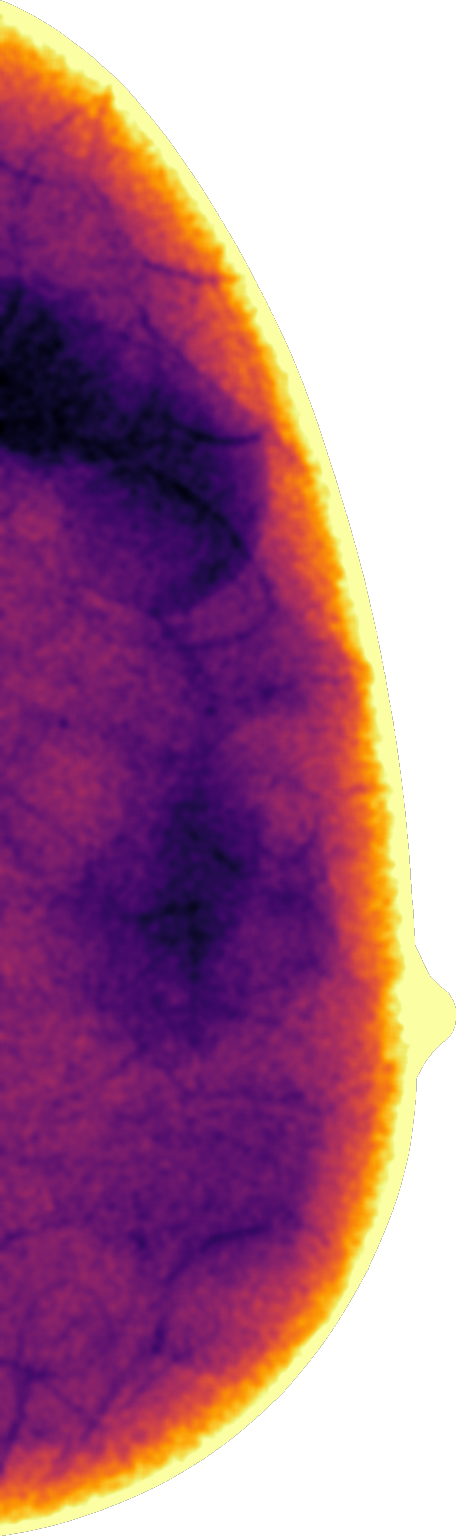}}
	\subfloat{\includegraphics[scale=0.46]{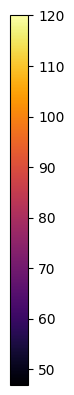}}
	\caption{\textcolor{reviewcolor}{Illustration of the SNR map of the anthropomorphic breast phantom images. (a) FD acquisition; (b) LD acquisition for a dose reduction factor of 75\%; restored images generated by the (c) MB method; (d) proposed network with the loss function MSE, (e) MAE, (f) SSIM and (g)-(j) PL1 to PL4, respectively. The maps were adjusted and clipped in the range of 47-120, based on the SNR of the FD image.}}
	\label{fig:snr_75}
\end{figure*}

\begin{figure*}[htb]
	\centering	
	\subfloat[FD]{\includegraphics[scale=0.115]{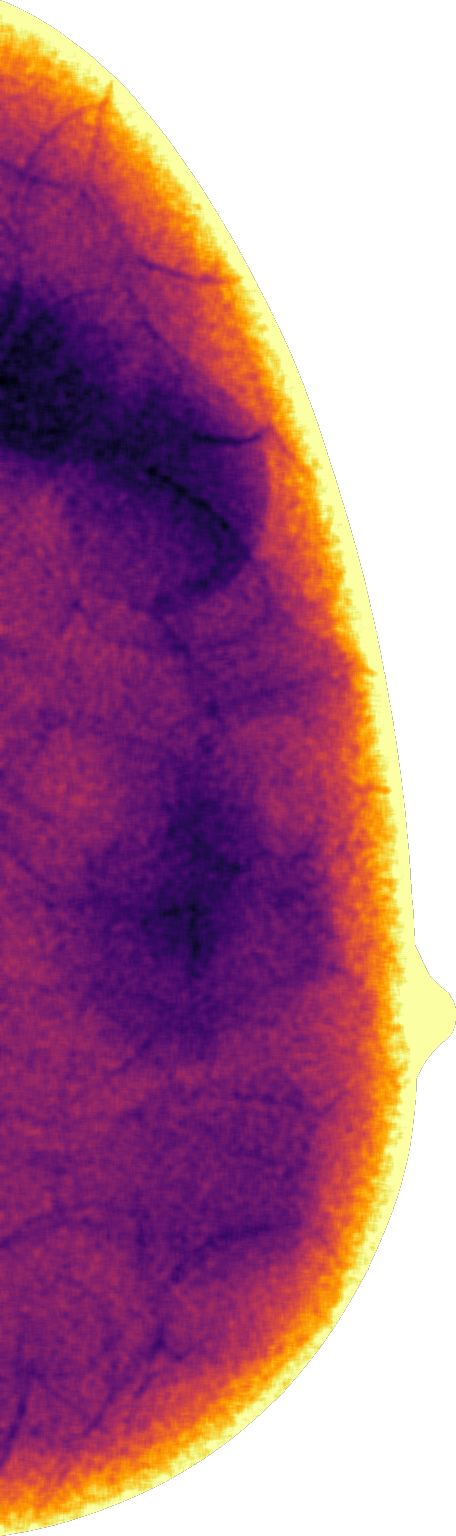}}
	\subfloat[LD]{\includegraphics[scale=0.115]{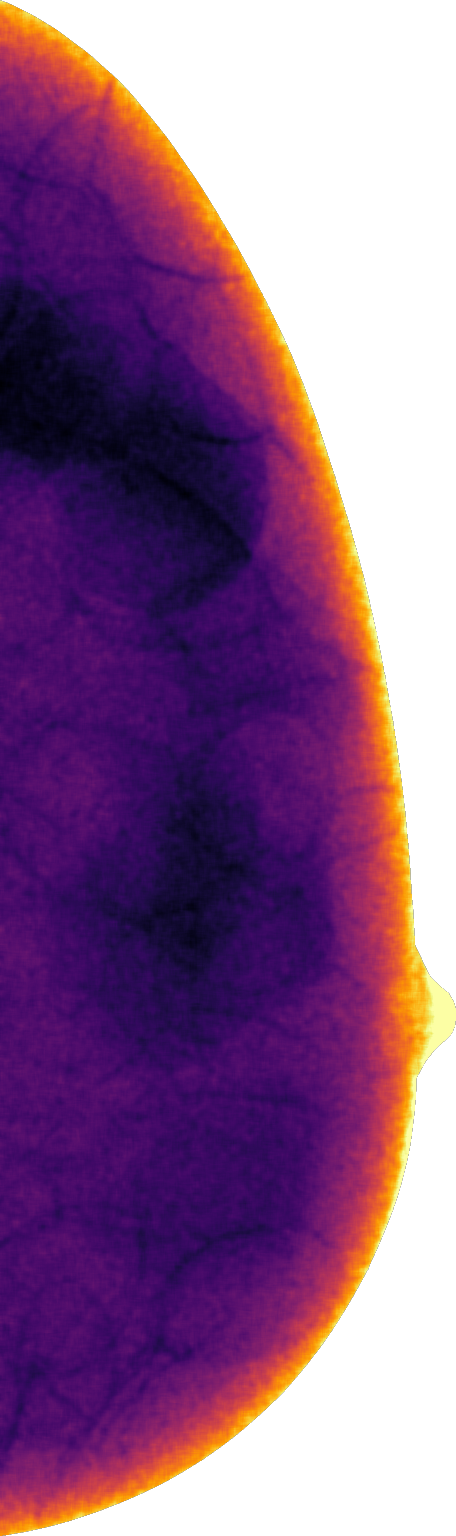}}
	\subfloat[MB]{\includegraphics[scale=0.115]{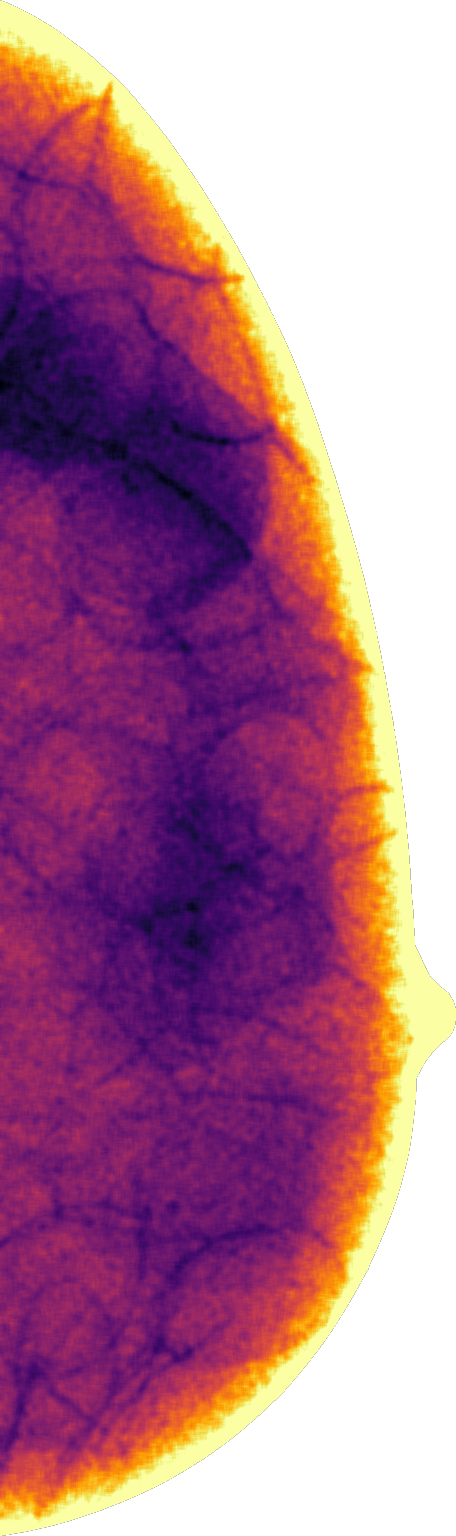}}
	\subfloat[$\mathcal{L}_{\mathrm{MSE}}$]{\includegraphics[scale=0.115]{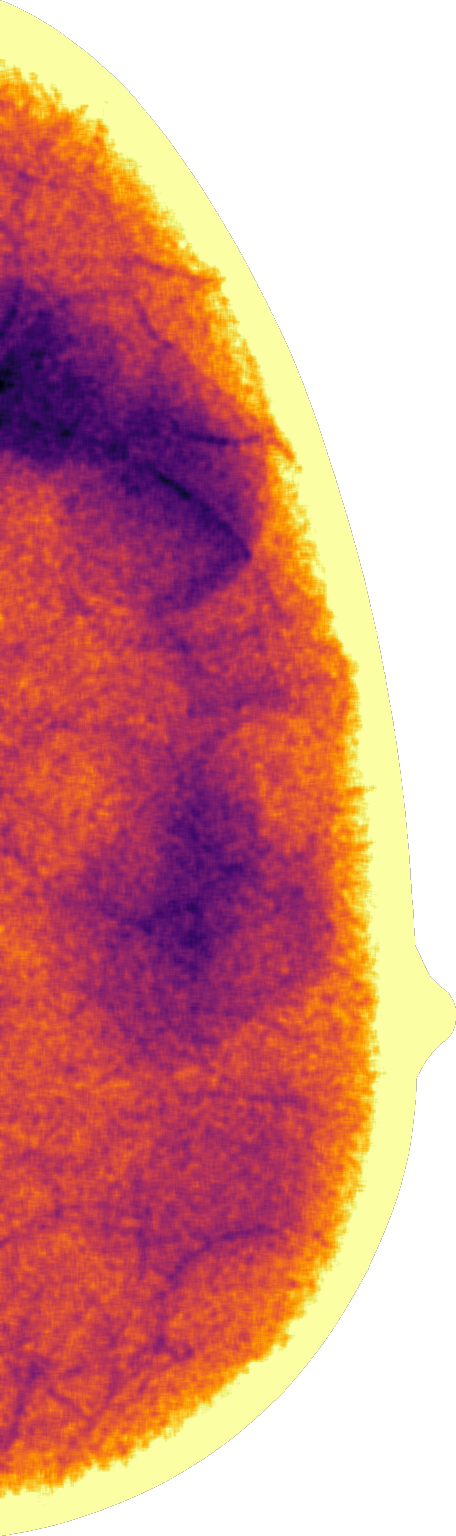}}
	\subfloat[$\mathcal{L}_{\mathrm{MAE}}$]{\includegraphics[scale=0.115]{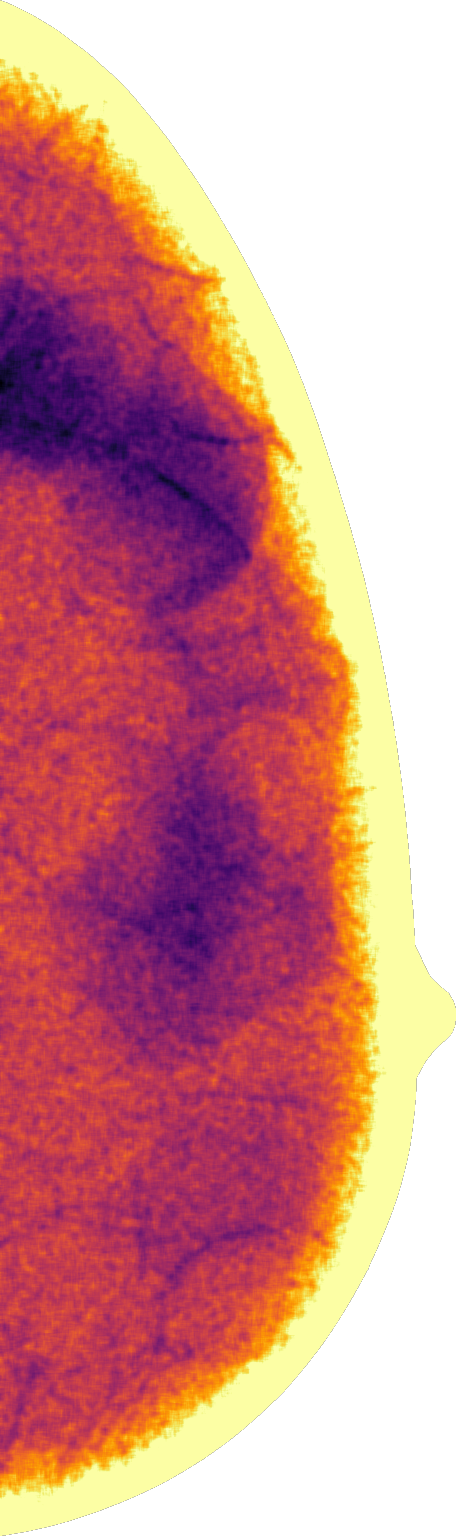}}
	\subfloat[$\mathcal{L}_{\mathrm{SSIM}}$]{\includegraphics[scale=0.115]{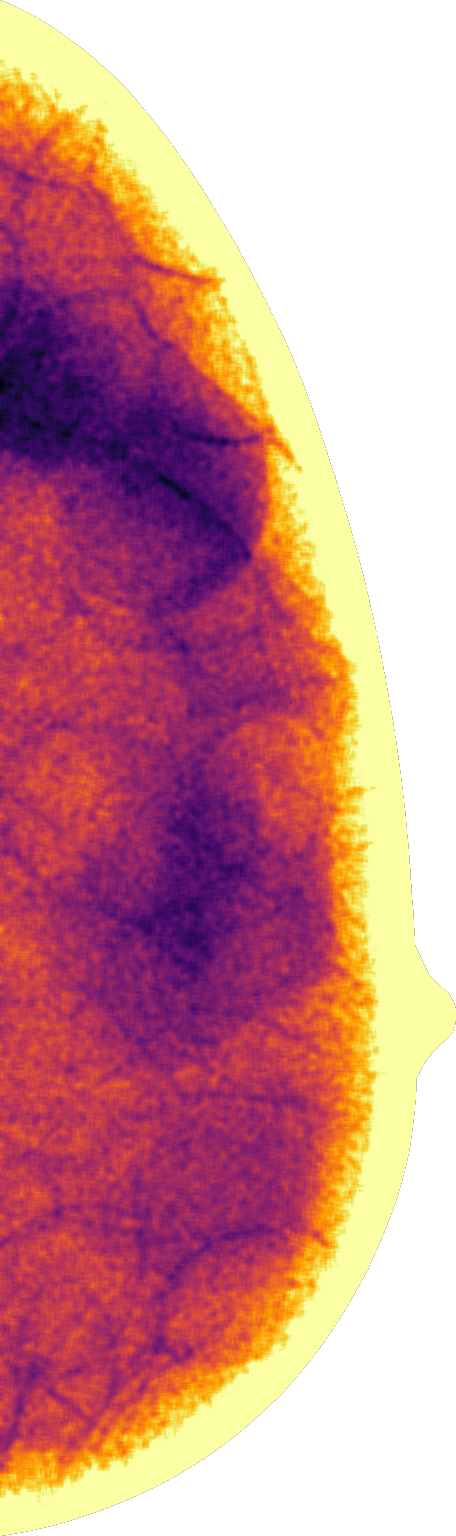}}
	\subfloat[$\mathcal{L}_{\mathrm{PL1}}$]{\includegraphics[scale=0.115]{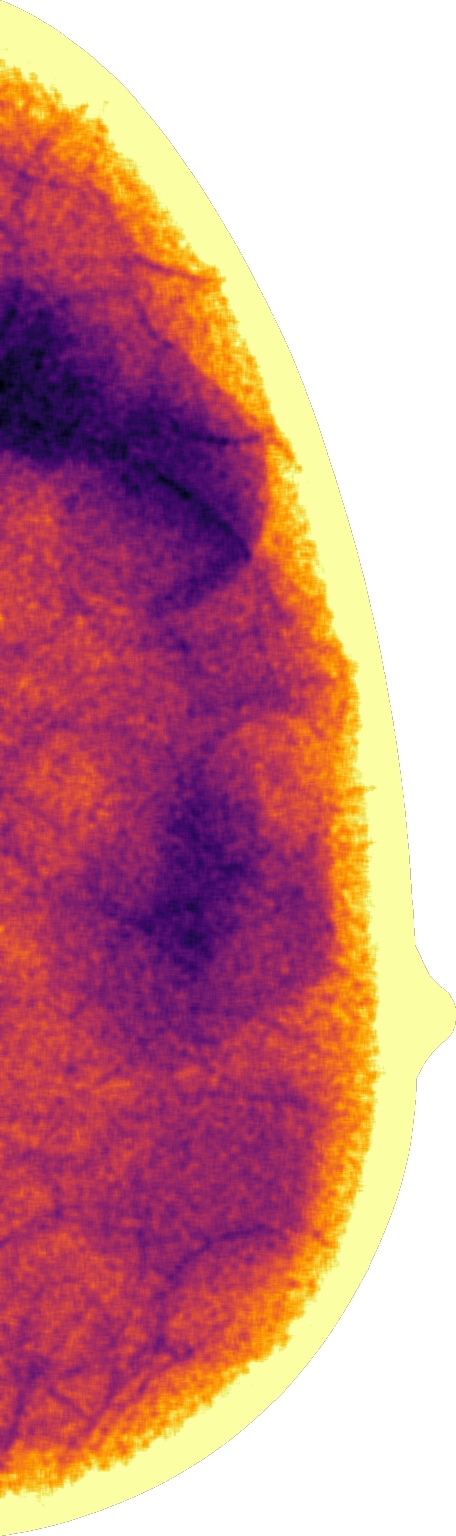}}
	\subfloat[$\mathcal{L}_{\mathrm{PL2}}$]{\includegraphics[scale=0.115]{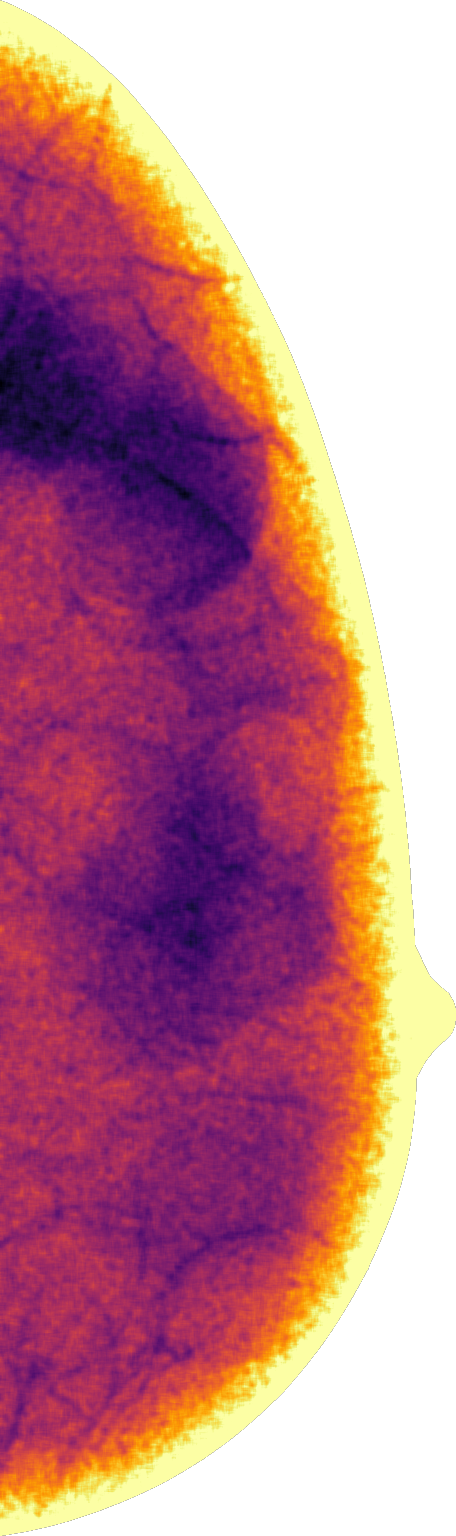}}
	\subfloat[$\mathcal{L}_{\mathrm{PL3}}$]{\includegraphics[scale=0.115]{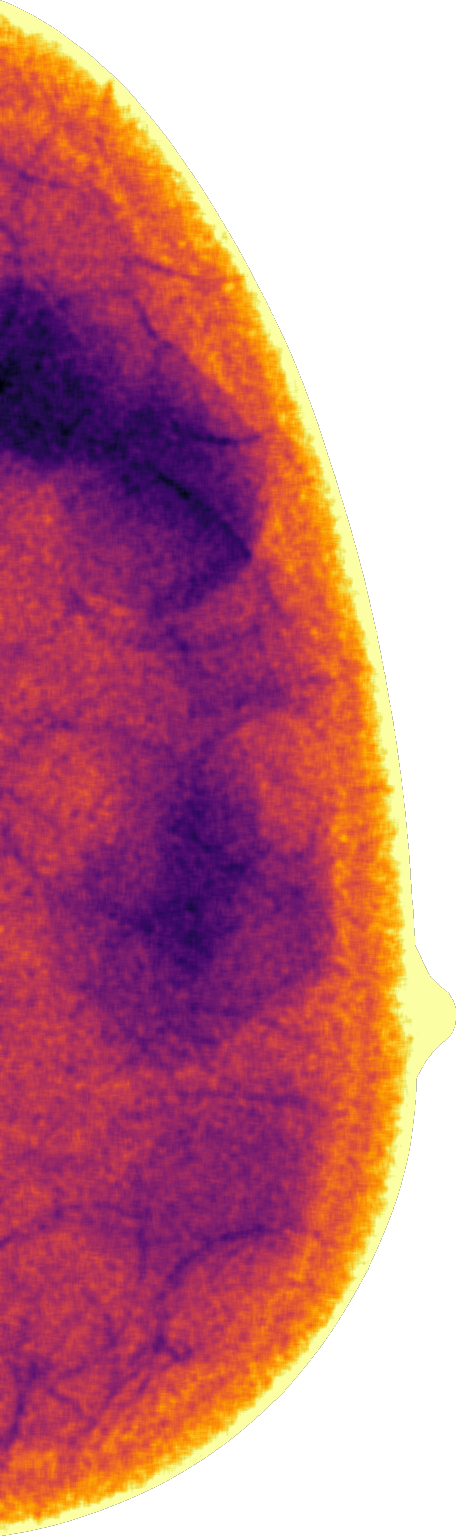}}
	\subfloat[$\mathcal{L}_{\mathrm{PL4}}$]{\includegraphics[scale=0.115]{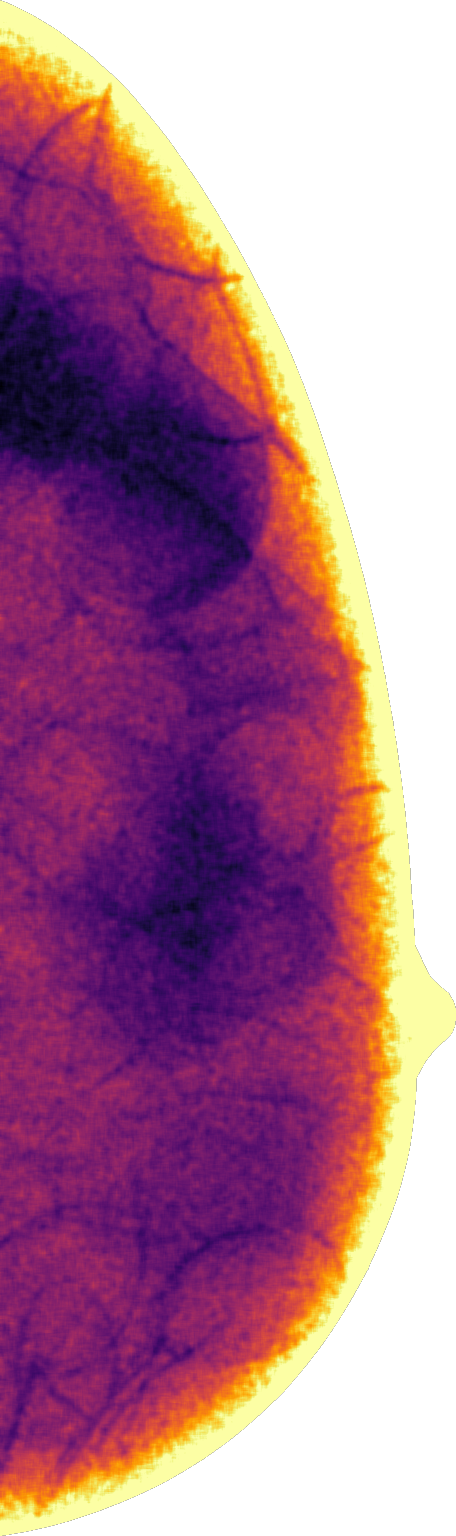}}
	\subfloat{\includegraphics[scale=0.46]{imgs/snr/colorbar.png}}
	\caption{\textcolor{reviewcolor}{Illustration of the SNR map of the anthropomorphic breast phantom images. (a) FD acquisition; (b) LD acquisition for a dose reduction factor of 50\%; restored images generated by the (c) MB method; (d) proposed network with the loss function MSE, (e) MAE, (f) SSIM and (g)-(j) PL1 to PL4, respectively. The maps were adjusted and clipped in the range of 47-120, based on the SNR of the FD image.}}
	\label{fig:snr_50}
\end{figure*}

Figs.~\ref{fig:phantom_75_imgCap6BR3D} and~\ref{fig:phantom_50_imgCap6BR3D} display a magnified ROI from the mammography image of the anthropomorphic breast phantom at dose reduction factors of 75\% and 50\%, respectively. In this case, it is important to note that the radiation dose reduction was performed on the equipment itself, \textcolor{newaddcolor}{changing the mAs with each acquisition}. The same discussion previously mentioned for the clinical images can be used here as the images have the same visual properties. From the results of the anthropomorphic phantom we can infer that the neural network is generalizing well even for cases with slightly different noise properties, \ie, even with a careful simulation as used in this work some discrepancies are expected between the simulated LD image and the actual LD image; and the trained model was able to overcome such small discrepancies. Also, it reinforces the fact that the network might be tested in real cases where the mammography is acquired at a reduced dose, even though it was trained with the simulated injection of noise.

\subsection{\textcolor{reviewcolor}{Quantitative Evaluation}}

\textcolor{reviewcolor}{Figs.~\ref{fig:snr_75} and~\ref{fig:snr_50} illustrate the SNR map inside the breast region of the anthropomorphic breast phantom image. Through a visual inspection, we can see that all the restorations were able to increase the SNR value from the LD. Also, the maps indicate that the restorations with MSE, MAE and SSIM achieved greatest values throughout the breast. Table~\ref{tab:SNR} presents the mean SNR value, demonstrating the aforementioned statements. The observance of higher SNR values for these loss functions is explained as they remove more noise compared with the other restorations, thus achieving a lower standard deviation.}

\textcolor{reviewcolor}{From Table~\ref{tab:SNR} it is observable that the MSE loss function yielded the highest SNR, even higher than the SNR of the FD image, thus indicating the best image quality. However, a quick inspection of Fig.~\ref{fig:phantom_50_imgCap6BR3D}~(d) shows that the MSE loss function results in significant signal smearing/blurring, especially if compared to the FD image in Fig.~\ref{fig:phantom_50_imgCap6BR3D}~(a). This emphasizes the need for a metric sensitive to signal smoothing and residual noise separately. To that end, we adopted the decomposition of the NMSE into $\mathcal{B}^2$ and $\mathcal{R}_{\mathcal{N}}$.}

\begin{table}[!t]
	\centering
	\caption{\textcolor{reviewcolor}{Mean SNR values for the anthropomorphic breast phantom images acquired at the FD and dose reduction factors of 75\% and 50\% (LD). Also, the results for the restored images generated by the MB method and for the proposed network with the loss function MSE, MAE, SSIM and PL1 to PL4. The confidence interval is displayed inside the brackets.}}
	\begin{threeparttable}
		\begin{tabular*}{1.0\linewidth}{@{\extracolsep{\fill}}lll} 
			\toprule
			& \multicolumn{1}{c}{75\%} & \multicolumn{1}{c}{50\%} \\ 
			\midrule
			LD                         & 69.18 [69.15, 69.20] & 56.41 [56.39, 56.44]  \\ 
			\hline
			DL-$\mathcal{L}_{\mathrm{MSE}}$ & 84.40 [84.38, 84.43] & 96.92 [96.89, 96.95]  \\
			DL-$\mathcal{L}_{\mathrm{MAE}}$ & 84.58 [84.55, 84.61] & 96.10 [96.07, 96.13]  \\
			DL-$\mathcal{L}_{\mathrm{SSIM}}$ & 84.51 [84.48, 84.54] & 95.58 [95.55, 95.61]  \\
			DL-$\mathcal{L}_{\mathrm{PL1}}$ & 84.10 [84.07, 84.13] & 94.51 [94.47, 94.54]  \\
			DL-$\mathcal{L}_{\mathrm{PL2}}$ & 82.41 [82.38, 82.43] & 86.89 [86.87, 86.92]  \\
			DL-$\mathcal{L}_{\mathrm{PL3}}$ & 79.30 [79.27, 79.32] & 84.50 [84.47, 84.52]  \\
			DL-$\mathcal{L}_{\mathrm{PL4}}$ & 78.45 [78.42, 78.48] & 80.39 [80.37, 80.42]  \\
			MB    & 78.39 [78.36, 78.42] & 77.53 [77.50, 77.56]  \\ 
			\hline
			FD                        & \multicolumn{2}{c}{78.11 [78.08, 78.13]}     \\
			\bottomrule
		\end{tabular*}
	\end{threeparttable}
	\label{tab:SNR}
\end{table}

\begin{table*}[!htb]
	\centering
	\caption{\textcolor{newaddcolor}{Quantitative analysis of the total MNSE, $\mathcal{B}^2$ and $\mathcal{R}_{\mathcal{N}}$ for the anthropomorphic breast phantom images acquired at the FD and dose reduction factor of 75\% (LD). Also, the results for the restored images generated by the MB method and for the proposed network with the loss function MSE, MAE, SSIM and PL1 to PL4. The confidence interval is displayed inside the brackets.}}
	\begin{threeparttable}
		\begin{tabular*}{0.8\textwidth}{@{\extracolsep{\fill}}lccc}
			\toprule
			& {Total MNSE(\%)} & {$\mathcal{R}_{\mathcal{N}}$(\%)} & {$\mathcal{B}^2$(\%)} \\ \midrule
			LD & 14.16 [14.08, 14.24] & 14.02 [14.00, 14.03] & 0.14 [0.14, 0.15] \\ \hline
			DL-$\mathcal{L}_{\mathrm{MSE}}$ & 9.71 [9.66, 9.77] & 9.37 [9.36, 9.38] & 0.34 [0.34, 0.35] \\
			DL-$\mathcal{L}_{\mathrm{MAE}}$ & 9.39 [9.33, 9.44] & 9.10 [9.09, 9.11] & 0.29 [0.28, 0.29] \\
			DL-$\mathcal{L}_{\mathrm{SSIM}}$ & 9.45  [9.40, 9.51] & 9.18 [9.17, 9.19] & 0.27 [0.27, 0.28] \\
			DL-$\mathcal{L}_{\mathrm{PL1}}$ & 9.59 [9.54, 9.65] & 9.34 [9.33, 9.35] & 0.25 [0.24, 0.26]\\
			DL-$\mathcal{L}_{\mathrm{PL2}}$     & 9.96 [9.91, 10.00] & 9.71 [9.70, 9.72] & 0.25 [0.24, 0.25] \\
			DL-$\mathcal{L}_{\mathrm{PL3}}$     & 10.97 [10.91, 11.00] & 10.57 [10.56, 10.60] & 0.40 [0.39, 0.40] \\
			DL-$\mathcal{L}_{\mathrm{PL4}}$     & 11.03 [10.97, 11.10] & 10.80 [10.79, 10.80] & 0.23 [0.22, 0.24] \\ 
			MB & 11.03 [10.97, 11.09] & 10.86 [10.85, 10.87] & 0.16 [0.16, 0.17] \\\hline
			FD & 10.58 [10.55, 10.62] & 10.49 [10.48, 10.50] & 0.09 [0.08, 0.09] \\
			\bottomrule
		\end{tabular*}
	\end{threeparttable}
	\label{tab:120}
\end{table*}

\begin{table*}[!htb]
	\centering
	\caption{\textcolor{newaddcolor}{Quantitative analysis of the total MNSE, $\mathcal{B}^2$ and $\mathcal{R}_{\mathcal{N}}$ for the anthropomorphic breast phantom images acquired at the FD and dose reduction factor of 50\% (LD). Also, the results for the restored images generated by the MB method and for the proposed network with the loss function MSE, MAE, SSIM and PL1 to PL4. The confidence interval is displayed inside the brackets.}}
	\begin{threeparttable}
		\begin{tabular*}{0.8\textwidth}{@{\extracolsep{\fill}}lccc}
			\toprule
			& {Total MNSE(\%)} & {$\mathcal{R}_{\mathcal{N}}$ (\%)} & {$\mathcal{B}^2$(\%)} \\ \midrule
			LD & 21.96 [21.86, 22.06] & 21.76 [21.74, 21.79] & 0.19 [0.18, 0.20] \\ \hline
			DL-$\mathcal{L}_{\mathrm{MSE}}$ & 8.39 [8.37, 8.42] & 7.49 [7.48, 7.50] & 0.90 [0.89, 0.91] \\
			DL-$\mathcal{L}_{\mathrm{MAE}}$ & 8.30 [8.28, 8.33] & 7.49 [7.48, 7.50] & 0.81 [0.80, 0.82] \\
			DL-$\mathcal{L}_{\mathrm{SSIM}}$ & 8.33 [8.30, 8.35] & 7.55 [7.54, 7.56] & 0.77 [0.77, 0.78] \\
			DL-$\mathcal{L}_{\mathrm{PL1}}$ & 8.46 [8.43, 8.50] & 7.69 [7.68, 7.69] & 0.78 [0.77, 0.79]\\
			DL-$\mathcal{L}_{\mathrm{PL2}}$ & 9.87 [9.84, 9.90] & 9.18 [9.17, 9.19] & 0.69 [0.68, 0.70]     \\
			DL-$\mathcal{L}_{\mathrm{PL3}}$ & 10.52 [10.44, 10.60] & 9.71 [9.70, 9.72] & 0.81 [0.80, 0.82]   \\
			DL-$\mathcal{L}_{\mathrm{PL4}}$ & 11.38 [11.35, 11.40] & 10.81 [10.80, 10.80] & 0.57 [0.56, 0.57]  \\ 
			MB & 12.00 [11.95, 12.05] & 11.65 [11.64, 11.66] & 0.35 [0.35, 0.36] \\\hline
			FD & 10.58 [10.55, 10.62] & 10.49 [10.48, 10.50] & 0.09 [0.08, 0.09] \\
			\bottomrule
		\end{tabular*}
	\end{threeparttable}
	\label{tab:80}
\end{table*}

\textcolor{reviewcolor}{As our primary goal is to restore the LD images to achieve the quality of the standard FD images, we desire a resulting image which has the overall characteristics of the standard FD. Thus, we seek a restoration method that yields an image with similar $\mathcal{R}_{\mathcal{N}}$ compared with the FD and as low $\mathcal{B}^2$ error as possible. This intuition comes from the fact that our goal is to generate restored images that have similar noise properties to the FD images, and also that we want to keep the underlying signal characteristics as close as the original image, as radiologists tend to dislike overly smoothed images.}

\textcolor{reviewcolor}{To this end, the total MNSE was measured against the pseudo-GT and decomposed into $\mathcal{B}^2$ and $\mathcal{R}_{\mathcal{N}}$ for the FD and for both the radiation dose reduction factors, considering all different loss functions. Tables~\ref{tab:120} and~\ref{tab:80} present the MNSE results for the dose reduction factors of 75\% and 50\%, respectively.}

Changing loss functions directly affects the behavior of the deep network in terms of signal preservation and noise suppression. As we can see in Tables~\ref{tab:120} and~\ref{tab:80}, the MSE loss function tends to decrease the $\mathcal{R}_{\mathcal{N}}$ values lower than the standard FD (goal), at the cost of excessively blurring the image, thus increasing the $\mathcal{B}^2$ error. Although this loss function is extensively used in most applications, this blurring behavior is well known in the literature~\citep{zhao2016loss}. The MAE loss function (${\ell_1}$), when compared to the MSE, leads to more image details preservation, observed by lower $\mathcal{B}^2$ values but has a strong denoising effect, noted in the lower $\mathcal{R}_{\mathcal{N}}$ values for the 75\% case. For the SSIM loss function case, the network performed better compared to the MSE and MAE, reporting sightly lower $\mathcal{B}^2$  at similar $\mathcal{R}_{\mathcal{N}}$.

The PL function brings an interesting case. There is a tendency to increasing image detail preservation coming from the PL1 to the PL4, where the $\mathcal{R}_{\mathcal{N}}$ slightly increases whereas the $\mathcal{B}^2$ decreases. This behavior is explained by the fact that the deeper neural network layers are responsible for general image characteristics, whereas the initial layers are responsible for local fine details of the image, \ie, primitive information like edges~\citep{erhan2009visualizing, gu2018recent,zhang2018interpretable}. When looking at the essential properties of the images, in the case of deep layers, the network tends to penalize errors on the underlying signal which causes a decrease in $\mathcal{B}^2$. The opposite happened with the initial layers, where they are trying to match the fine details, thus performing a more aggressive local denoising, decreasing the $\mathcal{R}_{\mathcal{N}}$ and increasing the $\mathcal{B}^2$ error. The previous discussion brings a consequence that the deeper in the network the loss function analysis is done, less aggressive the denoising and more image details are preserved overall.

It is important to note the great similarity that the neural network with the PL4 loss function has with the MB in terms of $\mathcal{R}_{\mathcal{N}}$. Also, this loss function has the lowest $\mathcal{B}^2$ error when compared with the other losses. Although the $\mathcal{B}^2$ is higher for the deep network compared to the mathematical model, the data-based approach benefits from the fact that it does not need to know any previous information about the equipment and its physics acquisition process. 

Finally, Table~\ref{tab:Time} demonstrates the average time spent by the proposed network and also by the MB to restore a single raw clinical image of size $4096\times3328$. Although the deep network takes a very long time for training, for example, roughly 7 hours with MSE and \textcolor{newaddcolor}{32} hours with PL4 (respectively the minimum and maximum training time for all loss functions), the restoration process has a processing time of the same order of magnitude as the MB method. Note that we do not intend to compare processing times, as the DL runs on a GPU under Python language while the MB runs on a CPU using a MATLAB code. However, both methods have room for code optimization as fast processing time is especially important for clinical use.

\begin{table}[!htb]
	\centering
	\caption{Average processing time to run a full restoration on a single raw clinical mammography.}
	\begin{threeparttable}
		\begin{tabular*}{0.8\linewidth}{@{\extracolsep{\fill}}lc}
			\toprule
			Method & Time (s)    \\
			\midrule
			DL-based & 9.0        \\
			MB   & 16.5     \\
			\bottomrule
		\end{tabular*}
	\end{threeparttable}
	\label{tab:Time}
\end{table}

\section{Conclusion}
\label{sec:Conclusion}

In this work, we investigated the impact of various loss functions on the quality of LD mammograms restored by deep networks. We also \textcolor{newaddcolor}{proposed} a modification of a known CNN architecture to \textcolor{newaddcolor}{evaluate} such loss functions. 

In terms of loss functions, the MSE and MAE had strong denoising properties yielding excessive smoothness in the restored image, while the PLs functions preserved images details as we go deeply in the VGG-16 network, \ie, the PL4 function preserved more details compared to the PL1. This behavior was observed both in the quantitative results and also in the visual \textcolor{reviewcolor}{analysis}. Furthermore, it is possible to note with both quantitative results and \textcolor{reviewcolor}{visually} the similarity between the PL4 function and the MB method. 

The fact that we used a physical anthropomorphic breast phantom to validate the proposed methodology reinforces that the neural network is able to restore real LD mammography images. This also implies that the network did not overfit on the training dataset and it is generalizing well for other images with slightly different noise properties.

\textcolor{reviewcolor}{With this work, we showed the potential of DNNs for DM image restoration and evaluated some well-known loss functions, presenting the strength and weaknesses of each one so we may choose which one is appropriated for each task. Also, with the new training strategy proposed, it is possible to use clinical images and the networks can learn and benefits from a great variability of data and their radiographic factors, as illustrated in Fig.~\ref{fig:exames_statistics}.} 

\textcolor{reviewcolor}{The limitations of this work are presented as follows. First, since we focus on the comparison among different loss functions, the comparison of different network architectures for the restoration is not considered in this article. As argued in~\cite{shan20183}, loss function is relatively more important than network architecture as the loss function has a direct impact on the image quality of the restored images. Second,  we did not consider the a combination of different loss functions in this paper. Although the combination of several loss functions could improve the results, it will bring extra balancing hyperparameters to be carefully tuned and exponentially more combinations; in case of combination of two loss functions, there will be 21 cases. Third, the network was not tested on real clinical images as there will no GT to evaluate the performance of different loss functions. Finally, we restricted the training to the dataset representing a local and specific woman population.}

For future works, real LD mammography images \textcolor{newaddcolor}{should} be tested through the network and cancer detectability evaluations \textcolor{newaddcolor}{should} be performed with radiologists to analyze the relevance of the proposed method in the clinical routine. Also, it is important to test the network with other datasets to see if the network is generalizing well for other populations as the dataset used in the current study is limited to certain characteristics of some population. 

\section*{Acknowledgments}

This work was supported in part by the National Council for Scientific and Technological Development (CNPq), by the S\~{a}o Paulo Research Foundation (FAPESP grant 2018/19888-5), by the \textit{Coordenação de Aperfeiçoamento de Pessoal de Nível Superior} (CAPES finance code 001) and National Institute of Health (R01EB026646, R01CA233888, R01CA237267, and R01HL151561).

The authors would like to thank Dr. Andrew D. A. Maidment, from the University of Pennsylvania, for making the anthropomorphic breast phantom images available for this work. The authors also would like to thank the Barretos Cancer Hospital, in particular the medical physicist Renato F. Caron, for providing the clinical images.

\bibliographystyle{unsrt}

\end{document}